\documentclass[11pt]{article}

\usepackage{comment}

\usepackage{a4wide}

\usepackage{mediabb}
\usepackage[dvipdfm]{graphicx}

\usepackage{epsf}
\usepackage{color}

\usepackage{amsmath}
\usepackage{amssymb}
\usepackage{latexsym}
\usepackage{slashed}
\usepackage{float}
\usepackage{multirow}

\DeclareMathOperator{\tr}{tr}

\DeclareMathOperator{\Br}{Br}

\newcommand{\MKK}{M_\text{KK}}
\newcommand{\fbinv}{\,\text{fb}^{-1}}
\newcommand{\GeV}{\,\text{GeV}}
\newcommand{\TeV}{\,\text{TeV}}
\newcommand{\shat}{\hat{s}}

\newcommand{\eps}{\epsilon}
\newcommand{\vep}{\varepsilon}

\newcommand{\A}{\mathcal{A}}
\newcommand{\B}{\mathcal{B}}
\newcommand{\D}{\mathcal{D}}
\newcommand{\E}{\mathcal{E}}
\newcommand{\F}{\mathcal{F}}
\newcommand{\G}{\mathcal{G}}
\renewcommand{\L}{\mathcal{L}}
\newcommand{\Q}{\mathcal{Q}}

\newcommand{\U}{\mathcal{U}}
\newcommand{\V}{\mathcal{V}}
\newcommand{\W}{\mathcal{W}}
\newcommand{\X}{\mathcal{X}}

\newcommand{\ol}{\ensuremath{\overline}}

\newcommand{\al}[1]{\begin{align}#1\end{align}}

\newcommand{\bp}{\begin{pmatrix}}
\newcommand{\ep}{\end{pmatrix}}
\newcommand{\bb}{\begin{bmatrix}}
\newcommand{\eb}{\end{bmatrix}}
\newcommand{\nn}{\nonumber\\}
\newcommand{\paren}[1]{\left(#1\right)}
\newcommand{\sqbr}[1]{\left[#1\right]}
\newcommand{\ab}[1]{\left|#1\right|}
\newcommand{\fn}[1]{\!\left(#1\right)}

\newcommand{\mc}{\mathcal}

\DeclareMathOperator{\im}{Im}
\DeclareMathOperator{\re}{Re}

\newcommand{\cred}{}
\newcommand{\cmagenta}{}
\newcommand{\skipped}{}
\newcommand{\skippedMagenta}{}

\newcommand{\Slash}[1]{{\ooalign{\hfil#1\hfil\crcr\raise.167ex\hbox{/}}}}
\newcommand{\beq}{\begin{equation}}
\newcommand{\eeq}{\end{equation}}
\newcommand{\pal}{\partial}

\begin{document}

\title{
Heavy Higgs at Tevatron and LHC\\
in Universal Extra Dimension Models
}
\author{
	\Large
	Kenji Nishiwaki,\thanks{
		E-mail: \tt \cred{nishiwaki@hri.res.in}
		}
	{}
	Kin-ya Oda,\thanks{
		E-mail: \tt odakin@phys.sci.osaka-u.ac.jp
		}\smallskip\\
	\Large
	Naoya Okuda,\thanks{
		E-mail: \tt okuda@het.phys.sci.osaka-u.ac.jp
		}
	{} and
	Ryoutaro Watanabe\thanks{
		E-mail: \tt ryoutaro@het.phys.sci.osaka-u.ac.jp
		}
	\medskip\\
	$^*$\it Department of Physics, Kobe University, Kobe 657-8501, Japan\smallskip\\
	\cred{$^*$\it Harish-Chandra Research Institute, Chhatnag Road, Jhusi, Allahabad 211 019, India} \smallskip\\
	$^\dagger{}^\ddag{}^\S$\it Department of Physics,  
	Osaka University,  Osaka 560-0043, Japan
	\smallskip\\
	}

\maketitle
\begin{abstract}\noindent
Universal Extra Dimension (UED) models tend to favor a distinctively heavier Higgs mass than in the Standard Model (SM) and its supersymmetric extensions when the Kaluza-Klein (KK) scale is not much higher than the electroweak one, which we call the weak scale UED, in order to cancel the KK top contributions to the $T$-parameter. Such a heavy Higgs, whose production through the gluon fusion process is enhanced by the KK top loops, is fairly model independent prediction of the weak scale UED models regardless of the brane-localized mass structure at the ultraviolet cutoff scale. We study its cleanest possible signature, the Higgs decay into a $Z$ boson pair and subsequently into four electrons and/or muons, in which all the four-momenta of the final states can be measured and both the $Z$ boson masses can be checked. 
\skipped
We have \skipped studied the Higgs mass 500\,GeV (and also 700\,GeV with $\sqrt{s}=14\TeV$) and have found that we can observe significant resonance with the integrated luminosity $10\fbinv$ for six dimensional UED models.
\end{abstract}
\vfill
\mbox{}\hfill KOBE-TH-11-05\\
\mbox{}\hfill OU-HET-721/2011
\newpage

\section{Introduction}
The Universal Extra Dimension (UED) scenario~\cite{Appelquist:2000nn}, in which all the Standard Model (SM) fields propagate in bulk of compactified extra dimension(s), is an attractive possibility whose simplest five dimensional realization on orbifold $S^1/Z_2$, the minimal UED model (mUED),  may account for the existence of the dark matter as the Lightest Kaluza-Klein Particle (LKP)~\cite{Servant:2002aq} and can give {a loose} gauge coupling unification at around 30\,TeV~\cite{Bhattacharyya:2006ym}.
See also Refs.~\cite{Kribs:2006mq,Hooper:2007qk} for review on mUED.
For the mUED, the latest analysis including the effects from second KK resonances gives the preferred Kaluza-Klein (KK) scale at around $\MKK\sim 1.3\,\text{TeV}$~\cite{Belanger:2010yx}. As is mentioned in~\cite{Servant:2002aq,Belanger:2010yx}, this result strongly depends on the brane-localized mass structure, which is assumed to be vanishing at the UV cutoff scale~\cite{Cheng:2002iz} in mUED.

One of the most important signature to establish the model would be the direct search of KK resonances at the CERN Large Hadron Collider (LHC). 
\nocite{Cheng:2002ab,Macesanu:2002db,Kazana:2007zz,Bhattacharyya:2009br,Choudhury:2009kz,Bhattacherjee:2010vm,Murayama:2011hj}
\nocite{Burdman:2006gy,Dobrescu:2007xf,Ghosh:2008ji,Ghosh:2008ix,Bertone:2009cb,Cacciapaglia:2011hx,Cacciapaglia:2011kz}
\nocite{Bhattacharyya:2005vm,Bhattacherjee:2005qe,Freitas:2007rh,Ghosh:2008dp,Battaglia:2005zf,Bhattacherjee:2008ik}
See Refs.~\cite{Cheng:2002ab}--\cite{Murayama:2011hj} for mUED and Refs.~\cite{Burdman:2006gy}--\cite{Cacciapaglia:2011kz} for 6D UED models.
\cred{We note that some of them also pertain to the International Linear Collider (ILC), see also \cred{Refs.~\cite{Bhattacharyya:2005vm}--\cite{Bhattacherjee:2008ik}.}} 
The LHC already puts a lower bound on the KK scale for mUED as $\MKK\gtrsim 500\,\text{GeV}$ at the 95\% CL from $M_{T2}$ analysis of cascade decay of first KK particles into the LKP~\cred{\cite{Murayama:2011hj,Tobioka:master}}.
\cred{It is noted that bounds on mUED~\cite{Haisch:2007vb} and $T^2/Z_4$ UED~\cite{Freitas:2008vh} from $b\to s\gamma$ processes claim $M_\text{KK}\gtrsim 600\GeV$ and 650\GeV, respectively.}
Again \cred{all the above bounds} strongly depend on the KK mass splitting \cred{and flavor mixing patterns} and hence on the boundary mass structure.\footnote{
A 95\% CL bound on the KK scale $\MKK>961\GeV$ is put on a ``UED'' model, assuming existence of large additional extra dimensions compactified with radius of order $\text{eV}^{-1}$, in which SM fields cannot propagate, so that the LKP decays into KK-gravitons~\cite{Aad:2011kz}. In this paper, we do not assume such additional large extra dimensions.
}
In particular, we cannot see a decay product unless there are enough mass splitting among the first KK modes so that it becomes sufficiently energetic.
\cred{In this paper, we present a complementary signal that is insensitive to such detailed boundary structure.}

In the SM, the electroweak data constrain the Higgs mass to be $M_H\lesssim 170\,\text{GeV}$ at the 95\% CL, see e.g.~\cred{\cite{Djouadi:2005gi,Riemann:2010zz}}. On the contrary, mUED  prefer heavier Higgs when the KK scale is not much higher than the electroweak scale $v_\text{EW}\simeq 246\GeV$, namely\cred{,
the KK scale should be $M_\text{KK}\gtrsim 800\GeV$ ($300\GeV\lesssim M_\text{KK}\lesssim 400\GeV$) at the 95\% CL for $M_H = 115$ (700)\,GeV}
~\cred{\cite{Appelquist:2002wb,Gogoladze:2006br,Baak:2011ze}}. We note that this is fairly model independent feature of a general UED model since KK top modes always contribute positively to the $T$-parameter and such an effect requires a heavy Higgs in order to cancel these KK top contributions by the ordinary negative $\log M_H$ dependence. 
In this paper, we call such a natural UED model without big mass splitting among electroweak, Higgs and KK scales: $M_H \sim M_\text{KK} \sim \mathcal{O}(10^2)$\,GeV, the {\it weak scale UED model}. Concretely, we will pick up the cases: $M_H = \skipped 330, 500$ and $700$\,GeV.
To summarize, existence of a heavy Higgs is a model independent prediction of the weak scale UED models in contrast to the relic abundance of the dark matter and the cascade decay signature of the KK particles that are dependent on the detailed boundary mass structure at the Ultra-Violet (UV) cutoff scale of the higher-dimensional gauge theory.

In five and six dimensional UED models, the Higgs production cross section via the gluon fusion process is enhanced by the KK top loops~\cite{Petriello:2002uu,Maru:2009cu,Nishiwaki:2011vi}. In this paper, we analyze its cleanest possible signature, the Higgs decay into a $Z$ boson pair and subsequently into four electrons and/or muons, in which all the four momenta of the final states can be measured and both the $Z$ boson masses can be checked.

In Section~\ref{UED_review_section}, we review the relevant part of all the known 5D and 6D UED models to the Higgs production process via the gluon fusion through the KK top loops. Considered models are the 5D mUED model on $S^1/Z_2$~\cite{Appelquist:2000nn}, the Dirichlet Higgs (DH) model on \cred{an} interval~\cite{Haba:2009pb,Nishiwaki:2010te}, the 6D $T^2$-based models on $T^2/Z_2$~\cite{Appelquist:2000nn}, $T^2/(Z_2\times Z_2')$~\cite{Mohapatra:2002ug}, $T^2/Z_4$~\cite{Dobrescu:2004zi,Burdman:2005sr}, $RP^2$~\cite{Cacciapaglia:2009pa}, and the 6D $S^2$-based models on Projective Sphere (PS)~\cite{Dohi:2010vc}, $S^2$ (see Section~\ref{S2_UED}), $S^2/Z_2$~\cite{Maru:2009wu}.\footnote{
In~\cite{Dohi:2010vc} the terminology ``real projective plane'' is employed for the compactified space, the sphere with its antipodal points being identified. In order to distinguish \cite{Dohi:2010vc} from \cite{Cacciapaglia:2009pa}, we call the former the Projective Sphere.
}
In Section~\ref{Higgs_production_and_decay_section}, we present the concrete computation of the process. The cross sections for the DH, $T^2/Z_2$, $T^2/(Z_2\times Z_2')$, $RP^2$, and $S^2$ UED models are newly obtained. We also review the estimation of the UV cutoff scale in the $T^2$-based geometry~\cite{Dienes:1998vg} and extend it to the $S^2$-based one.
In Section~\ref{numerical_results_section}, we show our numerical results.
The last section is for summary and discussions.
In Appendix~\ref{DH_calculation_in_appendix}, we present the relevant Feynman rules for our computation in the DH model. 
In Appendix~\ref{Details_of_renormalizationgroupanalysis} we explain our estimation of the UV cutoff scale for 6D UED models, based on the Renormalization Group Equation (RGE) analysis in the renormalizable KK picture.
In Appendix~\ref{Two_point_function_and_width}, we review the way to take into account the width in the amplitude and justify our approximation.

\section{Review on known 5D and 6D UED models}\label{UED_review_section}

In this section, we give a brief review on various UED models. Readers who are not interested in the details of these models may skip this section. In the first part of this section, we briefly review the 5D minimal UED model on $S^1/Z_2$~\cite{Appelquist:2000nn} and Dirichlet Higgs model on an interval~\cite{Haba:2009pb,Nishiwaki:2010te}.
The remaining of the section is devoted to an overview of various types of 6D UED models.

\subsection{5D UED models}

\subsubsection{Minimal UED model on $S^1/Z_2$}
First we review the 5D UED model~\cite{Appelquist:2000nn}. The matter contents of the model are the same as those of the SM, but they are living in the bulk of flat five-dimensional space, compactified on the orbifold $S^1/Z_2$. {The action $S$ is written down as}
\al{
S	&=	\int d^4x\int_{-\pi R}^{\pi R}dy\sqbr{
			\mc L_\text{bulk}+\delta(y)\mc L_0+\delta(y-\pi R)\mc L_{\pi R}}{.}
}
Usually when one say mUED model, it is implied that all the boundary masses are zero at the UV cutoff scale and are generated through radiative corrections~\cite{Cheng:2002iz}. Hereafter, when we call mUED model, we do not assume any boundary mass structure and concentrate on the signal that is independent of it. In particular, we do not include the constraints from the direct KK search~\cred{\cite{Murayama:2011hj,Tobioka:master}} and from the relic abundance of the LKP~\cite{Belanger:2010yx} that are dependent on the KK mass splitting pattern.

The $Z_2$ twist conditions on the bulk SM fields are put as
\al{
\B_\mu(x,-y)
	&=	\B_\mu(x,y),	&
\B_5(x,-y)
	&=	-\B_5(x,y),	\nn
\W_\mu(x,-y)
	&=	\W_\mu(x,y),	&
\W_5(x,-y)
	&=	-\W_5(x,y),	\nn
\G_\mu(x,-y)
	&=	\G_\mu(x,y),	&
\G_5(x,-y)
	&=	-\G_5(x,y),
	\label{mUEDboundarycondition_of_vectors}
}
\al{
L(x,-y)
	&=	\gamma^5L(x,y),	&
E(x,-y)
	&=	-\gamma^5E(x,y),\nn
Q(x,-y)
	&=	\gamma^5Q(x,y),	&
U(x,-y)
	&=	-\gamma^5U(x,y),\nn
&&
D(x,-y)
	&=	-\gamma^5D(x,y),
	\label{mUEDboundarycondition_of_fermions}
}
and
\al{
\Phi(x,-y)
	&=	\Phi(x,y),
	\label{mUEDboundarycondition_of_scalar}
}
where $x$ and $y$ \cred{($=x^5$)} denote four and extra dimensional coordinates\cred{, respectively}.\footnote{
We follow the metric and spinor conventions of~\cite{Weinberg:1995mt}.
}
We can see that the wanted zero modes remain after the twist~\eqref{mUEDboundarycondition_of_vectors}--\eqref{mUEDboundarycondition_of_scalar}.
There are fixed points of the $Z_2$ orbifolding at $y=0,\pi R$.
If the boundary Lagrangians at $y=0,\pi R$ are equal at the UV cutoff scale, there remains an additional accidental symmetry under the reflection ${\pi R\over2}-y\to{\pi R\over2}+y$, called the KK parity, which ensures the stability of the LKP and makes it a dark matter candidate.

The gauge and Yukawa interactions for the KK-top quarks, which we need for later calculation, are:
\al{
\mc L_\text{KK top}
&=	-  i g_{4s}
\sum_{n=1}^{\infty}
\bb \ol{t_1} & \ol{t_2} \eb^{(n)}
\gamma^{\mu} \G_{\mu}^{(0)}
\bb t_1 \\ t_2 \eb^{(n)}
\nn
&\quad	-{m_t\over v_{\text{EW}}} H^{(0)}
\sum_{n=1}^{\infty}
\bb \ol{t_1} & \ol{t_2} \eb^{(n)}
\bb \sin{2 \alpha^{(n)}} & -\gamma^5\cos{2 \alpha^{(n)}}  \\
\gamma^5\cos{2 \alpha^{(n)}}  & \sin{2 \alpha^{(n)}} \eb
\bb t_1 \\ t_2 \eb^{(n)},
	\label{L_KK_top}
}
where ${g_{4s} = g_s/\sqrt{2\pi R}}$ is a dimensionless 4D $SU(3)_C$ coupling constant 
and $v_{\text{EW}}\simeq246\GeV$ is the 4D Higgs vacuum expectation value
which appear after the KK expansion\cred{;}
$\G^{(0)} (H^{(0)})$ shows zero-mode gluon (zero-mode physical Higgs)\cred{;}
$t_1^{(n)}$ and $t_2^{(n)}$ are mass eigenstates of $n$-th KK top quarks
and each mixing angle $\alpha^{(n)}$ is determined to be $\cos{2 \alpha^{(n)}} = m_{(n)}/\sqrt{m_t^2 + m^2_{(n)}}$, $\sin{2 \alpha^{(n)}} = m_t/\sqrt{m_t^2 + m^2_{(n)}}$, with $m_{(n)}:=n/R$.
Each KK state is twofold degenerate and n-th KK top mass is
\al{
m_{t,(n)} = \sqrt{m_t^2 + m^2_{(n)}}.
}
KK tops give dominant contribution to the gluon fusion process due to their large Yukawa coupling to the Higgs.
\cred{We note that $\gamma^5$ is put in Eq.~\eqref{L_KK_top} merely to arrange the sign of both the KK masses positive.}

\subsubsection{Dirichlet Higgs (DH) model}
Dirichlet Higgs model is defined on an interval: $0 \leq y \leq \pi R$.
The action $S$ is as follows:
\al{
S	&=	\int d^4x\int_0^{\pi R}dy\sqbr{
			\mc L_\text{bulk}+\delta(y)\mc L_0+\delta(y-\pi R)\mc L_{\pi R}},
}
where $R$ is a radius of the extra spacial direction. 
{The structure of the bulk Lagrangian, covariant derivatives and field strength of gauge bosons are the same as that of the mUED model.
{There is no difference between the matter contents of this model and
those of the mUED model.}
As in the mUED model, we neglect the possible boundary interactions in this paper.
The zero-mode sector of the UED on an interval becomes the same as that of the mUED on the orbifold $S^1/Z_2$ when we choose the boundary conditions for the SM degrees of freedom $\Psi^N=\G_\mu,\W_\mu,\B_\mu;L_L,Q_L;E_R,U_R,D_R$ to be Neumann (at $y=0$ and $\pi R$):
\al{
\partial_{\cred 5}\Psi^N(x,0)
	=	\partial_{\cred 5}\Psi^N(x,\pi R)
	=	0
}
and for other non-SM modes $\Psi^D=\B_{\cred 5},\W_{\cred 5};L_R,Q_R;E_L,U_L,D_L$ to be Dirichlet:
\al{
\Psi^D(x,0)=\Psi^D(x,\pi R)=0.
}
{We note that mode functions with Dirichlet and Neumann boundary conditions
are not orthogonal to each other, unlike the orbifolding on $S^1/Z_2$.\footnote{
In other words the {KK mass-squared operator} $\partial_{\cred 5}^2$ is not hermitian in this setup, though the kinetic term is still positive definite.
}
Kinetic terms turn out to be diagonal even though the expansion is not orthonormal. We can explicitly check that the non-orthogonality does not lead to extra mixing for spinors
even after the EWSB because non-orthogonal terms drop out due to the 4D-chirality.

If we had put the Neumann condition on the Higgs $\Phi$, we would get exactly the same zero-mode sector as in the mUED model on $S^1/Z_2$.
In the Drichlet Higgs model on interval, the EWSB is caused by a non-zero Dirichlet boundary condition on the $SU(2)_W$-doublet Higgs field~\cite{Haba:2009pb,Nishiwaki:2010te}.
We assume that the KK-parity is respected by the boundary conditions on the Higgs field too. 
The advantage of the Dirichlet EWSB is that we do not need to assume the negative mass-squared in the bulk Lagrangian nor the quartic coupling which is a higher dimensional operator in 5D. Throughout this paper, we consider the minimal case: $\V(\Phi) = 0$.
We list the necessary Feynman rules in Appendix~\ref{DH_calculation_in_appendix}.

\subsection{6D UED models based on $T^2$}


We consider a gauge theory on six-dimensional spacetime $M^4 \times T^2$, which is a direct product of the four-dimensional Minkowski spacetime $M^4$ and two-torus $T^2$: $0 \leq y \leq 2 \pi R_y, 0 \leq z \leq 2\pi R_z$.
We assume that the two radii of $T^2$ have the same value $R = R_y = R_z$ for
simplicity.\footnote{
{In the $T^2/Z_4$ orbifold case, the condition $R_y = R_z$ is imposed by the consistency with the $Z_4$ discrete symmetry.}
See also~\cite{Lim:2009pj} for a realization of CP violation from the complex structure of $T^2/Z_4$, which appears in 4D effective interactions after KK decomposition.}

When we use 6D Weyl spinor for 6D UED model construction, there is a constraint on the choice of 6D chiralities. 
The origin of this constraint is the cancellation of 6D gravitational and SU(2)$_{L}$ global anomalies that cannot be removed by use of the Green-Schwarz mechanism. This constraint requires the number of matter generation to be (multiple of) three~\cite{Dobrescu:2001ae}.
A suitable choice of the 6D chirality for a single matter generation is as follows:
\al{
({Q}_+ , {U}_- , {D}_- ; {L}_+ , {E}_- , {N}_-),
\label{eq:Weyl_choice}
}
where the $\pm$ suffixes represent 6D chirality of each field.
Number of d.o.f.\ of 6D Weyl fermion is~4, the same as that of a 4D Dirac fermion. Therefore we can construct 6D UED models on $T^2$ following the orbifolding method of the 5D UED model. We have several options for the orbifolding to realize the SM chiral fermions in the zero mode sector of \eqref{eq:Weyl_choice}. Let us review them in turn.
The range for KK summation is listed in Table~\ref{range_of_mandn}.

\subsubsection{Orbifold ${T^2/Z_2}$, $T^2/(Z_2 \times Z'_2)$, and $T^2/Z_4$}

\begin{table}[t]
\begin{center}
\begin{tabular}{|c|c|c|}
\hline 
type of orbifolding & identification & fixed points $(y_i,z_i)$ \\
\hline 
$T^2/Z_2$ & $(y,z) \sim (-y,-z)$ & $(0,0),\ (\pi R, 0),\ (0,\pi R),\ (\pi R, \pi R)$\\
$T^2/(Z_2 \times Z'_2)$
& $(y,z) \sim (-y,z)$ and $(y,z) \sim (y,-z)$ & $(0,0),\ (\pi R, 0),\ (0,\pi R),\ (\pi R, \pi R)$\\
$T^2/Z_4$ & $(y,z) \sim (-z,y)$ & $(0,0),\ (\pi R,\pi R)$ \\
\hline
\end{tabular}
\caption{Fixed points which stem from each identification.}
\label{fixedpoints_of_T2}
\end{center}
\end{table}

We consider ${T^2/Z_2}$~\cite{Appelquist:2000nn}, $T^2/(Z_2 \times Z'_2)$~\cite{Mohapatra:2002ug},
and $T^2/Z_4$~\cite{Dobrescu:2004zi,Burdman:2005sr} orbifolds.
Let us write down the action
\al{
S	&=	\int d^4x\int_{-\pi R}^{\pi R}dy\int_{-\pi R}^{\pi R}dz
			\sqbr{
			\mc L_\text{bulk}(x,y,z) + \sum_{\vec{y}\  \in\ \vec{y}_{i} } \delta(\vec{y}_{i})\mc L_{\vec{y}_{i}}(x)},
}
where $\vec{y}_{i} = (y_i,z_i)$ are orbifold fixed points.
We note that terms localized at the fixed points are induced at quantum level even if we assume that they are vanishing at tree level~\cite{Cheng:2002iz,Ponton:2005kx,Azatov:2007fa}.
%
For the orbifold we consider in this paper, the projections are:
\al{
(y,z)
	&\sim (-y,-z)  &
	&\text{for $T^2/Z_2$}\label{form_of_T2Z2}, \\
(y,z) \sim (-y,z)\
	&\text{and}\ (y,z) \sim (y,-z)	&
	&\text{for $T^2/(Z_ 2\times Z'_2)$}, \\
(y,z)
	&\sim (-z,y)  &
	&\text{for $T^2/Z_4$}.
} 
See also Table.~\ref{fixedpoints_of_T2}.
In each case, we can choose a suitable boundary condition of 6D Weyl fermions,
whose exact forms are not discussed in this paper,
to generate 4D Weyl fermions
at the zero modes.

The bulk Lagrangian, covariant derivatives and field strengths of gauge bosons are essentially the same as that of the 5D mUED model 
except 
for the structure of spinors.
For 6D Weyl fermions,
the kinetic and Yukawa terms are 
\al{
\mc L_\text{kinetic}
	&=	- \ol{Q_+}\Gamma^M\D_MQ_+
		- \ol{U_-}\Gamma^M\D_MU_-
		- \ol{D_-}\Gamma^M\D_MD_- \nn
	&\quad
		- \ol{L_+}\Gamma^M\D_ML_+
		- \ol{E_-}\Gamma^M\D_ME_-
		- \ol{N_-}\Gamma^M\D_MN_-,\\
\mc L_\text{Yukawa}
	&=	
		- \lambda_{U} \ol{U_-} \paren{Q_+\cdot\Phi}
		- \lambda_{D} \paren{\ol{Q_+} \Phi} D_-	
		- \lambda_{E} \paren{\ol{L_+} \Phi} E_-
		+ \text{h.c.},
}
where contraction of $SU(2)$ indices are understood.\footnote{
We leave the neutrino sector untouched since it is irrelevant for the Higgs signal considered in this paper.
}
%
Resultant interactions relevant for our discussion are
\al{
\mc L_\text{KK top}
	&=	-  i g_{4s}
		\sum_{(m,n)}^{\infty}
		\bb \ol{t_1} & \ol{t_2} \eb^{(m,n)}
		\gamma^{\mu} \G_{\mu}^{(0)}
		\bb t_1 \\ t_2 \eb^{(m,n)}\nn
	&\quad
		- \frac{m_t}{v_{\text{EW}}} H^{(0)}
	\sum_{(m,n)}^{\infty}
	\bb \ol{t_1} & \ol{t_2} \eb^{(m,n)}
	\bb \sin{2 \alpha^{(m,n)}} & -\gamma^5\cos{2 \alpha^{(m,n)}}  \\
	\gamma^5\cos{2 \alpha^{(m,n)}}  & \sin{2 \alpha^{(m,n)}} \eb
	\bb t_1 \\ t_2 \eb^{(m,n)},
	\label{T2orbifoldinteractions}
}
where ${g_{4s} = g_s/(2 \pi R)}$ is the dimensionless 4D $SU(3)_C$ coupling constant and $v_{\text{EW}}\simeq 246\GeV$ is the 4D Higgs vev, 
$\G^{(0)} (H^{(0)})$ shows zero-mode gluon (zero-mode physical Higgs),
and $t_1^{(m,n)}, t_2^{(m,n)}$ are mass eigenstates of $(m,n)$-th KK top quarks.
Again we only consider the KK top quark loops since contributions from other flavors are suppressed by the small Yukawa coupling.
Each mixing angle $\alpha^{(m,n)}$ is determined to be
$\cos{2 \alpha^{(m,n)}} = m_{(m,n)}/\sqrt{m_t^2 + m^2_{(m,n)}}$,\ $\sin{2 \alpha^{(m,n)}} = m_t/\sqrt{m_t^2 + m^2_{(m,n)}}$.
Each KK state is twofold degenerate and $(m,n)$-th KK top mass is
\al{
m_{t,(m,n)} = \sqrt{m_t^2 + m^2_{(m,n)}},
}
with 
\al{
m_{(m,n)} := \frac{\sqrt{m^2+n^2}}{R}.
}
{It should be mentioned that the difference from the mUED case
appears only in the form of KK mass and 
the number of d.o.f.\ in each KK level
{when we consider the gluon fusion process}.}
We adopt m(n) as the y(z)-directional KK index, whose parameter region 
is determined by the way of the orbifolding.
This information has a great influence on 
the enhancement of the Higgs production through the gluon fusion.

\begin{table}[t]
\begin{center}
\begin{tabular}{|c|c|}
\hline 
type of orbifolding & range of $(m,n)$ \\
\hline 
$T^2/Z_2$ & {$m+n \geq 1,$ or $m=-n \geq 1$} \\
$T^2/(Z_2 \times Z'_2)$ & $0 \leq m < \infty, \ 0 \leq n < \infty;\ (m,n) \not=(0,0)$ \\
$T^2/Z_4$ & $1 \leq m < \infty, \ 0 \leq n < \infty$ \\
\hline
\end{tabular}
\caption{The range of the parameter $(m,n)$ except the zero mode case $(m,n)=(0,0)$
in each case of the orbifolding.\label{range_of_mandn}}
\end{center}
\end{table}%

\subsubsection{Real Projective Plane $(RP^2)$}

We can construct a UED model on a non-orientable geometry:
Real Projective Plane ($RP^2$)~\cite{Cacciapaglia:2009pa}.
$RP^2$ is {defined} by two types of identifications: a $\pi$-rotation $r$ and a glide $g$: 
\al{
r:(y,z) \sim (-y,-z),\quad g:(y,z) \sim (y+\pi R, -z + \pi R).
\label{pirotation_and_glide}
}
The system is invariant under each manipulation in Eq.~(\ref{pirotation_and_glide}). Note that the shifts $y\sim y+2\pi R$ and $z\sim z+2\pi R$ can be obtained as different combinations of $r$ and $g$, respectively.
Note also that no fixed point exists globally in this background geometry.
Under $r$ and $g$, Weyl fermions transforms as
\al{
r: \Psi_{{\pm}}(x;-y,-z) &= p_r \Gamma_r \Psi_{{\pm}}(x;y,z),
\quad \Gamma_r = \cred{i} 
\Gamma^\cred{5} \Gamma^\cred{6} \Gamma^\cred{7}, 
\label{r_fermion_condition} \\
g: \tilde{\Psi}_{{\pm}}(x;y+\pi R,-z+\pi R) &=
p_g \Gamma_g  \Psi_{{\mp}}(x;y,z),
\quad \Gamma_g = {\Gamma^\cred{6} \Gamma^\cred{7}},
\label{g_fermion_condition} 
}
where \cred{$\Gamma^7$ is the 6D chirality operator and} $p_r,p_g$ ($Z_2$-parities) can take the value $\pm 1$\skipped.
The $\tilde{\Psi}_{{\pm}}$ is what we call the ``mirror" fermion. 
{Eq.}~(\ref{r_fermion_condition}) has the same form with that of the $T^2/Z_2$ orbifold condition for 6D fermion.
An essential point of this model is that the condition~\eqref{g_fermion_condition} does not generate a 4D Weyl fermion in the zero mode sector.
In other words, the 6D chirality of both sides of {Eq.}~(\ref{g_fermion_condition}) are different from each other.
{This means that we have to introduce new fermions $\tilde{\Psi}_{{\pm}}$ which have opposite 6D chirality and the same SM quantum number compared to each corresponding field $\Psi_{{\mp}}$}.
Concretely, ``mirror" fermions:
\al{
{\Q}_- , {\U}_+ , {\D}_+ ; {\L}_- , {\E}_+ , \mathcal{N}_+,
\label{mirrorfermions}
}
are identified with $\{ {Q}_+ , {U}_- , {D}_- ; {L}_+ , {E}_- , {N}_- \}$, respectively.
The choice of 6D chiralities in Eq.~(\ref{mirrorfermions}) obeys the condition for realizing the 6D anomaly cancellation which we have argued before.

The bulk Lagrangian is the same as that of the $T^2$-based models
using orbifold except for the existence of the mirror fermions:
\al{
\mc L_\text{kinetic}
  &= \frac{1}{2} \Big[ - \ol{Q_+}\Gamma^M\D_MQ_+
		- \ol{U_-}\Gamma^M\D_MU_-
		- \ol{D_-}\Gamma^M\D_MD_-\notag \\
	&\phantom{= \frac{1}{2} \Big[\ } - \ol{L_+}\Gamma^M\D_ML_+
		- \ol{E_-}\Gamma^M\D_ME_-
		- \ol{N_-}\Gamma^M\D_MN_-\notag \\
	&\phantom{= \frac{1}{2} \Big[\ } - \ol{\Q_-}\Gamma^M\D_M\Q_-
		- \ol{\U_+}\Gamma^M\D_M\U_+
		- \ol{\D_+}\Gamma^M\D_M\D_+\notag \\
	&\phantom{= \frac{1}{2} \Big[\ } - \ol{\L_-}\Gamma^M\D_M\L_-
		- \ol{\E_+}\Gamma^M\D_M\E_+
		- \ol{\mathcal{N}_+}\Gamma^M\D_M \mathcal{N}_+\Big],
		\label{RP2_kinetic}\\	
\mc L_\text{Yukawa}
	&=	{1\over2}\Big[
			- \lambda_{U} \ol{U_-} \paren{Q_+\cdot\Phi}
			- \lambda_{D} \paren{\ol{Q_+} \Phi} D_-	
			- \lambda_{E} \paren{\ol{L_+} \Phi} E_-
		\nn
	&\phantom{=	{1\over2}\big[\ }
			- \lambda_{U} \ol{\U_+} \paren{\Q_-\cdot\Phi}
			- \lambda_{D} \paren{\ol{\Q_-} \Phi} \D_+
			- \lambda_{E} \paren{\ol{\L_-} \Phi} \E_+
			+ \text{h.c.}
			\Big],
}
where we introduce the $``1/2"$ factors for later convenience.
The neutrino sector is again left untouched as it is irrelevant for our discussion.
By use of Eq.~\eqref{g_fermion_condition}, we can erase all the mirror fermions and obtain the ordinary form of Lagrangian same as $T^2$ cases.
The form of $\pi$-rotation $r$ given in Eq.~\eqref{r_fermion_condition} is the same as that of $Z_2$ orbifolding in Eq.~(\ref{form_of_T2Z2}).
Therefore, the interactions of $RP^2$ model, needed to calculate the gluon fusion process, take the same form as that of $T^2/Z_2$ one given in Eq.~(\ref{T2orbifoldinteractions}).

\subsection{6D UED models based on $S^2$}

Let us review UED models based on the $S^2$ compactification.
We span the extra dimension by the zenith and azimuthal angles $\theta$ and $\phi$, respectively.
The two-sphere $S^2$ has a positive curvature and 
to stabilize the {radius $R$}, we introduce an extra $U(1)_X$ gauge field which has a monopole-like classical configuration~\cite{RandjbarDaemi:1982hi} 
\al{
[\X^c_{\phi}(x^{\mu},\theta,\phi)]^{{N} \atop {S}}
	&=	{ n \over 2g_X }(\cos{\theta} \mp 1),	&
(\text{other components})
	&=	0,	\label{eq:monopoleconfig}
}
where
	the superscript $c$ denotes the classical configuration,
	$g_X$ is the {6D} $U(1)_X$ gauge coupling,
	the integer $n$ is the (negative) monopole charge, and 
	the superscripts $N$ and $S$ indicate that the field is given in north (involving the $\theta = 0$ point) and
south (involving the $\theta = \pi$ point) charts, respectively. 
The {$U(1)_X$} transition function 
from the north to the south chart is given by
\beq
[\X_M(x^{\mu},\theta,\phi)]^{S} = [\X_M(x^{\mu},\theta,\phi)]^{N} + \frac{1}{g_X} \pal_M \alpha(x^{\mu},
\theta,\phi)
\label{eq:gaugepatchtransf}
\eeq
with $\alpha(x^{\mu},\theta,\phi) = n \phi$.
Because of the monopole-like configuration, the radius of $S^2$ is stabilized spontaneously at
\beq
R^2 = \left( {n \over 2 g_X M_{\ast}^2} \right)^2,
\label{eq:radius}
\eeq
where $M_{\ast}$ is the 6D Planck scale. 

{We mention that any} 6D field ${\Xi}$ on $S^2$ is KK expanded by use of the spin-weighted spherical harmonics 
${}_s Y_{jm}(\theta,\phi)$ as follows: 
\beq
{\Xi}(x,\theta,\phi)^{N \atop S} = \sum_{j = |s|}^{\infty} \sum_{m=-j}^{j} {\xi}^{(j,m)}(x) f_{{\Xi}}^{(j,m)}(\theta,\phi)
^{N \atop S}, \quad
f_{{\Xi}}^{(j,m)}(\theta,\phi)
^{N \atop S} := {{}_s Y_{jm}(\theta,\phi) e^{\pm is \phi} \over R},
\eeq
where {$\xi^{(j,m)}$ is the $(j,m)$-th expanded 4D field,
$f^{(j,m)}_{\Xi}$ is the corresponding mode function and}
$s$ is the spin weight of the field ${\Xi}$.
The spin-weighted spherical harmonics ${}_s Y_{jm}(\theta,\phi)$ matches the orthonormal condition as 
\al{
\int_{0}^{2\pi} d\phi \int_{-1}^{1} d\cos{\theta} \ \overline{{}_s Y_{jm}(\theta,\phi)}\,
{}_s Y_{j'm'}(\theta,\phi) = \delta_{jj'} \delta_{mm'}.
}
A spin weight of fermion is closely related to its $U(1)_X$ charge.
When we assign $U(1)_X$ charges of 6D Weyl fermions $\Psi_{\pm}$ as $q_{\Psi_{\pm}}$,
the corresponding spin weights of 4D Weyl fermions $\{ \psi_{+ {L \atop R}},\psi_{- {L \atop R}} \}$ {are} given as follows in our convention:
\beq
s_{+ {L \atop R}} = 
	{nq_{\Psi_+} \pm 1 \over 2 }, \quad
s_{- {L \atop R}} = 
	{nq_{\Psi_-} \mp 1 \over 2 }.
\eeq
Note that if a 6D Weyl fermion takes a spin weight $s=0$,
a $j=0$ mode appears as a 4D Weyl fermion with vanishing KK mass.
This means that we can get chiral SM fermions without orbifolding in the case of $S^2$.
When we take the values: 
\al{
(s_{+R},s_{+L},s_{-R},s_{-L}) = (0,1,1,0),
}
we can create the same situation as in the $T^2$-based models discussed before.
{The} spin weight of {the} 4D-vector component of {a} 6D gauge boson is always $s=0$ and then there is a zero mode that can be identified as the SM gauge boson. On the other hand, extra dimensional components of the 6D gauge boson are expanded by the $|s|=1$ spin-weighted spherical harmonics and has no zero-mode.

In our configuration, any $(j,m)$-th KK mode has the KK mass:
\beq
m_{(j,m)} = \frac{\sqrt{j(j+1)}}{R}.
\label{eq:KKmass_S2}
\eeq
An important point is that the form of the above KK mass is independent of the index of $m$.
This means that there are $2j+1$ degenerate modes for each $j$.
Note that the lightest KK mode has the mass $\sqrt{2}/R$.

As discussed above, the 4D-vector component of a 6D gauge boson has a zero mode.
This is the case for the extra $U(1)_X$ gauge boson too.
Phenomenologically the existence of an extra $U(1)$ interaction, under which SM fields are charged, is problematic~\cite{Dohi:2010vc}. In the following, let us see how to get rid of this massless $U(1)_X$ vector.

\subsubsection{Projective Sphere (PS)}

We can construct a UED model compactified on the Projective Sphere $(PS)$, a sphere $S^2$ with its antipodal points being identified by $(\theta,\phi) \sim (\pi - \theta , \phi + \pi)$~\cite{Dohi:2010vc}. In the UED model based on PS, the 6D action takes a different form from that of the 6D orbifold UED models.
One of the remarkable points of this model is that
there is no fixed point on the background geometry PS. 
As in the $RP^2$ model, we introduce ``mirror" 6D Weyl fermions:
\al{
{\Q}_{-} , {\U}_{+} , {\D}_{+} ; {\L}_{-} , {\E}_{+} , \mathcal{N}_{+}
}
which have opposite 6D chirality and opposite
SM and $U(1)_X$ charges, compared to the fields
$\{ {Q}_{+} , {U}_{-} , {D}_{-} ; {L}_{+} , {E}_{-} , {N}_{-} \}$.
Because of the existence of mirror fermions the kinetic term takes the same form as in the $RP^2$ model~\eqref{RP2_kinetic} and the Yukawa interaction is modified to
\al{
\mc L_\text{Yukawa}
	&=	{1\over2}\Big[
			- \lambda_{U} \ol{U_-} \paren{Q_+\cdot\Phi}
			- \lambda_{D} \paren{\ol{Q_+} \Phi} D_-	
			- \lambda_{E} \paren{\ol{L_+} \Phi} E_-
		\nn
	&\phantom{=	{1\over2}\Big[}
			- \lambda_{U}^* \ol{\U_+} \paren{\Q_-\cdot\Phi}
			- \lambda_{D}^* \paren{\ol{\Q_-} \Phi} \D_+
			- \lambda_{E}^* \paren{\ol{\L_-} \Phi} \E_+
			+ \text{h.c.}
			\Big].
}
Like the $RP^2$ case which we have discussed before, we introduce the $``1/2"$ factors for a later convenience.
The covariant derivatives in this model are given as
\al{
{\D_M} &= \partial_M+ig_s\G_M^aT^a_s+ig\W_M^aT^a+ig_Y\B_MY  \notag\\
&
\hspace{80mm} (\text{for $\Phi$}), \\
{\D_M} &= \partial_M+ig_s\G_M^aT^a_s+ig\W_M^aT^a+ig_Y\B_MY
+ig_X q_{\Psi} (\X_M^c + \X_M) + \Omega_M \notag \\
&
\hspace{80mm} (\text{for ${Q}_{+} , {U}_{-} , {D}_{-} ; {L}_{+} , {E}_{-} , {N}_{-}$}),
\label{eq:ordinarycov} \\
{\D_M} &= \partial_M+ig_s\G_M^a[-T^a_s]^{\text{T}}+ig\W_M^a[-T^a]^{\text{T}}+ig_Y\B_M[-Y]
+ig_X q_{\Psi} (\X_M^c + \X_M) + \Omega_M \notag \\
& \hspace{80mm}
(\text{for ${\Q}_{-} , {\U}_{+} , {\D}_{+} ; {\L}_{-} , {\E}_{+} , \mathcal{N}_{+}$}),
\label{eq:mirrorcov}
}
where $\Omega_M$ is the spin connection.
The covariant derivative of Higgs is the same as that in the $S^2/Z_2$ case, 
but there is a difference between those of fermions and these ``mirror" fermions.
We discuss these points shortly below.

{As we mentioned before,}
projective sphere is a non-orientable manifold and has no fixed point.
Let us consider the 6D $P$ and $CP$ transformations.
Under the antipodal projection,
\beq
\left\{
\begin{array}{lcc}
\X_{\mu}(x,\pi - \theta,\phi + \pi)^{N \atop S} & = & \X_{\mu}^C(x,\theta,\phi)^{S \atop N} , \\
\X_{\theta}(x,\pi - \theta,\phi + \pi)^{N \atop S}  &= & -\X_{\theta}^C(x,\theta,\phi)^{S \atop N}, \\
\{ \X_{\phi}^c , \X_{\phi} \}(x,\pi - \theta,\phi + \pi)^{N \atop S}  & = 
&\{ (\X_{\phi}^c)^C , \X_{\phi}^C \}(x,\theta,\phi)^{S \atop N},
\end{array}
\right.
\label{eq:idRP2X}
\eeq
where the superscript $C$ denotes the 6D C transformation. Recall that the superscript $c$ denotes the classical configuration. These conditions leave the monopole-like configuration invariant under the antipodal identification and projects out the unwanted $U(1)_X$ 4D-vector zero mode.
In contrast, identification of a SM gauge boson $\A^{(i)}_M$ should be done by another condition since we want the corresponding 4D-vector zero mode,
where $i$ shows the type of gauge group.
We adopt the 6D $P$ transformation and those identifications are written as
\beq
\left\{
\begin{array}{lcc}
\A^{(i)}_{\mu}(x,\pi - \theta,\phi + \pi)^{N \atop S} & = & \A^{(i)}_{\mu}(x,\theta,\phi)^{S \atop N} , \\
\A^{(i)}_{\theta}(x,\pi - \theta,\phi + \pi)^{N \atop S}  &= & -\A^{(i)}_{\theta}(x,\theta,\phi)^{S \atop N}, \\
\A^{(i)}_{\phi}(x,\pi - \theta,\phi + \pi)^{N \atop S}  & = 
& \A^{(i)}_{\phi}(x,\theta,\phi)^{S \atop N},
\end{array}
\right.
\label{eq:idRP2A}
\eeq
where it is evident that the zero mode of ${\A_{\mu}^{(i)}}$ survives.
We also identify Higgs with the 6D $P$ transformation to obtain its zero mode:
\beq
\Phi(x,\pi - \theta,\phi + \pi)^{N \atop S} = \Phi(x,\theta,\phi)^{S \atop N}.
\label{eq:idRP2Higgs}
\eeq

Finally, we discuss the identification of 6D Weyl fermions.
Since 6D Weyl fermions have $U(1)_X$ charge and interact with {the} $U(1)_X$ gauge boson,
they should be identified by the 6D $CP$ transformation.
The specific form {of the 6D $CP$ transformation}, for example {in} the case of $U_-$, is as follows:
\beq
\mathcal{U}_{+}(x,\pi - \theta,\phi + \pi)^{N \atop S} =
P {U}_{-}^C (x,\theta,\phi)^{S \atop N},
\label{eq:idRP2fermions}
\eeq
where the matter field $U_-$ is identified to the mirror $\mathcal{U}_{+}$.
We decide the forms of covariant derivatives~\eqref{eq:ordinarycov} and \eqref{eq:mirrorcov} on the criterion of invariance of the action under the 6D $CP$ transformation.
Using the identification conditions~(\ref{eq:idRP2X})--(\ref{eq:idRP2fermions}), we can see that the mirror fermions drop out of the action 
after the identifications 
and eventually we obtain the usual type of UED model action.
%
This can be interpreted that all the modes of mirror fermions $\{ {\Q}_{-} , {\U}_{+} , {\D}_{+} ; {\L}_{-} , {\E}_{+} , \mathcal{N}_{+} \}$ are erased and no mode of $\{ {Q}_{+} , {U}_{-} , {D}_{-} ; {L}_{+} , {E}_{-} , {N}_{-} \}$ is projected out.
The interaction terms which we need for calculation in this model is the same as those in Eq.~(\ref{T2orbifoldinteractions}). Only difference is the number of degenerate top KK modes in each $j$-level.


\subsubsection{$S^2$ UED with a Stueckelberg Field ($S^2$)}\label{S2_UED}
As a solution to the massless $U(1)_X$ problem, we can simply give a Stueckelberg mass~\cite{Stueckelberg:1900zz,Ogievetskii:1962zz}, see also~\cite{Ruegg:2003ps,Kors:2005uz} for reviews, to the $U(1)_X$ field.
We can make the unwanted $U(1)_X$ 4D-vector zero mode to be massive while preserving the classical monopole structure in Eq.~(\ref{eq:monopoleconfig}).
This way, we can formulate a UED model on $S^2$ with no field identification.
Let us call this simple model the $S^2$ UED model.
In the $S^2$ UED model, the matter contents, bulk Lagrangian, definition of
field strengths and covariant derivatives and the configuration of the classical $U(1)_X$ field are the same as the PS model after removing the mirror fermions, 
except for the Stueckelberg field part.
In contrast to the $S^2/Z_2$ orbifold below, there are no fixed point nor a localized Lagrangian anywhere on $S^2$, as in the case of PS UED. 
The d.o.f.\ of KK fermions has no difference between the $S^2$ UED and PS one  (after the antipodal projection).
There is no need for an additional computation; all we have to do is to borrow the PS result as a whole when we are only interested in the gluon fusion process.

\subsubsection{$S^2/Z_2$ orbifold}
Although the above $S^2$ UED model with a Stueckelberg $U(1)_X$ mass is already phenomenologically viable, we may further perform a $Z_2$ orbifolding on it~\cite{Maru:2009wu}.\footnote{
This extra $Z_2$ cannot project out the $U(1)_X$ gauge field. In~\cite{Maru:2009wu}, the $U(1)_X$ is assumed to be broken by an anomaly. Since we need a classical configuration of the $U(1)_X$, it would be theoretically preferable to break it by a tiny Stueckelberg mass.
}
On {this} orbifold, the point ${(\theta,\phi)}$ is identified with $(\pi - \theta , - \phi)$. The 6D action $S$ is as follows:
\al{
\hspace{-12mm}
S	&=	\int d^4x\int_0^{\pi}d\theta \int_0^{2\pi}d\phi \sqrt{-g}
			\sqbr{
			\mc L_\text{bulk}(x,y,z) +  \delta\paren{\theta - {\pi \over 2 }}
			\delta\paren{\phi} \mc L_{(\pi/2,0)}(x)
			+  \delta\paren{\theta - {\pi \over 2 }}
			\delta\paren{\phi - \pi} \mc L_{(\pi/2,\pi)}(x)},
			\label{eq:6DactionofS2Z2}
}
where $\sqrt{-g} = R^2 \sin{\theta}$.
This system has two fixed points of {the} $Z_2$ symmetry at $(\theta , \phi) = (\frac{\pi}{2},0) , (\frac{\pi}{2},\pi)$ and
we describe the localized terms with $\mc L_{(\pi/2,0)}, \mc L_{(\pi/2,\pi)}$,
respectively.
Like the $T^2$-case, we do not discuss those parts in this paper.

We can easily construct mode functions of $S^2/Z_2$ $f_{s,t}^{(j,m)}(\theta,\phi)$ with spin weight $s$
in both north and south charts
following the general prescription~\cite{Georgi:2000ks} as follows:
\beq
f_{s,t}^{(j,m)}(\theta,\phi)^{N \atop S} =
\left\{
\begin{array}{ll}
\displaystyle \frac{1}{2R}
\left[ {}_s Y_{jm}(\theta,\phi) + (-1)^{j-s} {}_s Y_{j-m}(\theta,\phi) \right] e^{\pm is \phi} & 
\text{for} \ t=+1 \\
\displaystyle \frac{1}{2R}
\left[ {}_s Y_{jm}(\theta,\phi) - (-1)^{j-s} {}_s Y_{j-m}(\theta,\phi) \right] e^{\pm is \phi} & 
\text{for} \ t=-1
\end{array}
\right. ,
\eeq
where $t=\pm 1$ is {the} $Z_2$ parity.
These mode functions have the property that
$f_{s,t=\pm 1}^{(j,m)}(\pi - \theta, - \phi)^{N \atop S} = \pm f_{s,t=\pm 1}^{(j,m)}(\theta,\phi)^{S \atop N}$.
To realize {the} $Z_2$ symmetry, we identify a field at $(\theta,\phi)$ in north chart with
the same field at $(\pi - \theta , - \phi)$ in south chart.

The range of the summation over $m$ shrinks from $[-j,j]$ to $[0,j]$ after {the} $Z_2$ identification.
Under the transformation of $(\theta,\phi) \rightarrow (\theta,\phi + \pi)$,
mode functions behave as 
\beq
f_{s=0,t=+1}^{(j,m)}(\theta,\phi + \pi)^{N \atop S} = (-1)^{m} f_{s=0,t=+1}^{(j,m)}(\theta,\phi)^{N \atop S},
\quad
f_{s=\pm 1,t=-1}^{(j,m)}(\theta,\phi + \pi)^{N \atop S} = -(-1)^m f_{s=\pm 1,t=-1}^{(j,m)}(\theta,\phi)^{N \atop S}.
\eeq
After some fields redefinition,
we can find that each KK field has a KK parity $(-1)^m$, which is a remnant of the KK angular momentum conservation. 

We focus on the $m=0$ modes of each $j$ level.
When we see the concrete forms of mode functions in $m=0$, which are
\begin{align}
f_{s=0,t=+1}^{(j,m=0)}(\theta,\phi)^{N \atop S} &= \frac{1}{2R} (1 + (-1)^j) \cdot
{}_0 Y_{j0}(\theta,\phi), \\
f_{s=+1,t=-1}^{(j,m=0)}(\theta,\phi)^{N \atop S} &= \frac{1}{2R} (1 + (-1)^j) \cdot
{}_1 Y_{j0}(\theta,\phi) e^{\pm i \phi}, \\
f_{s=-1,t=-1}^{(j,m=0)}(\theta,\phi)^{N \atop S} &= \frac{1}{2R} (1 + (-1)^j) \cdot
{}_{-1} Y_{j0}(\theta,\phi) e^{\mp i \phi},
\end{align}
{where} we find that $m=0$ modes appear only in the case of even $j$. 
Then degeneracy of KK masses is
\beq
\begin{array}{cl}
j+1 & \text{for} \quad j : \text{even}, \\
j & \text{for}  \quad j : \text{odd},
\end{array}
\eeq
since $m$ runs from $0$ to $j$.
Again, this mode counting is the only important point when computing the enhancement of the Higgs production via gluon fusion process.
%
%
We do not discuss the form of interactions which we need for calculating 
the gluon fusion process because there is essentially no difference from the $T^2$ case~(\ref{T2orbifoldinteractions}).

\section{Higgs production and decay into four leptons in UED models}\label{Higgs_production_and_decay_section}


\begin{figure}[t]
\centering
\includegraphics[width=10em, bb= 0 0 114 124,clip]{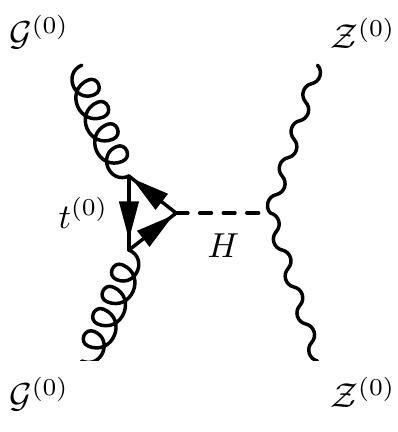}
\caption{Feynman diagram which describes the dominant contribution to the gluon fusion Higgs production process and the subsequent decay to 2Z.}
\label{SMgluonfusiondiagram}
\end{figure}

In the SM, the cross section for the Leading Order (LO) one-loop Higgs production via {the} gluon fusion process and its subsequent decay into {a} $Z$ boson pair: $gg\to H\to ZZ$ is given by~\cite{Georgi:1977gs}.
The \cmagenta{LO} parton-level cross section shown in Fig.~\ref{SMgluonfusiondiagram} is:
\al{
\hat\sigma^\text{SM}_{gg\to H\to ZZ}
	&=	{\alpha_{{4s}}^2\over256\pi^3}\paren{m_Z\over v_\text{EW}}^4
		\sqbr{
			1+{\paren{\shat-2m_Z^2}^2\over 8m_Z^4}
			}
		{\shat\over\paren{\shat-M_H^2}^2+ 
			\Delta^2
			}
		\sqrt{1-{4m_Z^2\over\shat}}\,
		\ab{   {I\fn{\cred{\shat}} }}^2,
		\label{SM_gg_H_ZZ}
}
where
$m_W, m_Z, m_t$, and $M_H$ are respectively the $W$, $Z$, top quark, and Higgs boson masses,
$\alpha_{4s} = {g_{4s}^2/4\pi}$ is the $4D$ QCD gauge coupling,
$\hat{s}$ is the center-of-mass-energy-squared of the scattering partons,
we employ the normalization for the Higgs vev: $v_\text{EW}^2=1/\sqrt{2}G_F\simeq\paren{246\,\text{GeV}}^2$,
and the loop functions are defined as
\al{
I(\lambda)
	&=	\cred{-}2\lambda+\lambda(\cred{1-4\lambda})
		\int_0^1{dx\over x}\ln\sqbr{{x(x-1)\over\lambda}+1-i\epsilon},\\
\tilde{I}(\lambda)
	&=	\skipped\lambda
		\int_0^1{dx\over x}\ln\sqbr{{x(x-1)\over\lambda}+1-i\epsilon}.
}
Explicit result of the integral is
\al{
\int_0^1{dx\over x}\ln\sqbr{{x(x-1)\over\lambda}+1-i\epsilon}
	&=	\begin{cases}
		\displaystyle -2\sqbr{\arcsin{1\over\sqrt{4\lambda}}}^2
			&	\text{(for $\lambda\geq{1\over4}$)},\\
		\displaystyle {1\over2}\sqbr{
			\ln{1+\sqrt{1-4\lambda}\over1-\sqrt{1-4\lambda}}
			-i\pi
			}^2
			&	\text{(for $\lambda<{1\over4}$)}.
		\end{cases}
}
We have also defined  $\tilde I$ for later use for the Dirichlet Higgs model. 
In Eq.~\eqref{SM_gg_H_ZZ}, we have taken into account the total decay width of the Higgs in its propagator:
\al{
\Delta = M_H\Gamma_H.
	\label{Delta_replacement}
}
In the current analysis, we take into account the Higgs {decay} into $W$, $Z$ and top quark pairs, which are dominant when we consider the heavy SM Higgs boson: $M_H \geq 2m_W$. 
Explicit form is shown in Eq.~\eqref{tree_width} in appendix.
Note that we take into account only the top quark loop in the SM cross section~\eqref{SM_gg_H_ZZ}, given by the diagram shown in Fig.~\ref{SMgluonfusiondiagram}, since the Yukawa coupling to others are negligible compared to the top one. We have also ignored the contributions from the sub-leading box diagrams~\cite{Glover:1988rg}. See Appendix~\ref{Two_point_function_and_width} for further discussion on how to take into account the width.

\subsection{Gluon fusion process in UED models}
In this paper, we consider the KK-top loop contributions to the gluon fusion process in several UED models, namely,
5D UED model on $S^1/Z_2$ (mUED)~\cite{Appelquist:2000nn}, Dirichlet Higgs {(DH)}~\cite{Haba:2009pb}, 6D UED model on $T^2/Z_2~\cite{Appelquist:2000nn},\ T^2/Z_4~\cite{Dobrescu:2004zi,Burdman:2005sr},\ T^2/(Z_2 \times Z'_2)~\cite{Mohapatra:2002ug}$, {Real Projective Plane (${RP}^2$)~\cite{Cacciapaglia:2009pa}, $S^2/Z_2$~\cite{Maru:2009wu}, Projective Sphere (PS)~\cite{Dohi:2010vc} and {$S^2$ with a Stueckelberg Field~($S^2$)}.
We have given a brief review {on} these models in the previous section.
The contribution from KK-top loops to the gluon fusion process is analogous to that of the top-loop in SM and the difference resides only in the loop function. 
Relevant Feynman diagram is shown in Figs.~\ref{UEDgluonfusiondiagrams} and \ref{effectivevertex}.
The effective vertex, which is represented by the lined blob in the diagram, includes
the contributions to the gluon fusion from the zero mode top quark and the KK top quarks.
Since the zero mode sector of UED model regenerates the SM configuration,
the \cred{result} of the former contribution is the same as that of the SM in Eq.~(\ref{SM_gg_H_ZZ}). The forms of the latter contribution will be shown soon later.
We note that the letter ``$H$"  in Figs.~\ref{UEDgluonfusiondiagrams} and \ref{effectivevertex} shows the zero mode physical Higgs boson 
in all the cases except for the DH model where $H$ stands for the first KK Higgs.


\begin{figure}[t]
\centering
\includegraphics[width=10em, bb= 0 0 117 122, clip]{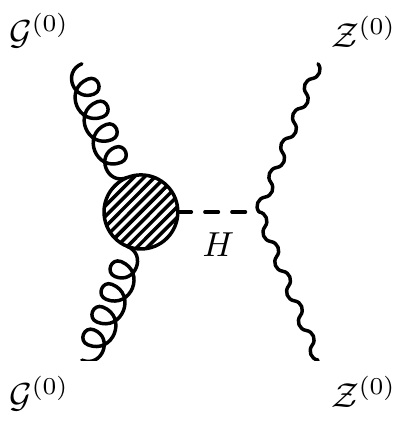}
\caption{A schematic description of the dominant contribution to the gluon fusion Higgs production process and the subsequent decay to 2Z. The lined blob indicates the effective vertex.}
\label{UEDgluonfusiondiagrams}
\end{figure}

\begin{figure}[t]
\centering
\includegraphics[width=25em, bb= 0 0 361 122,clip]{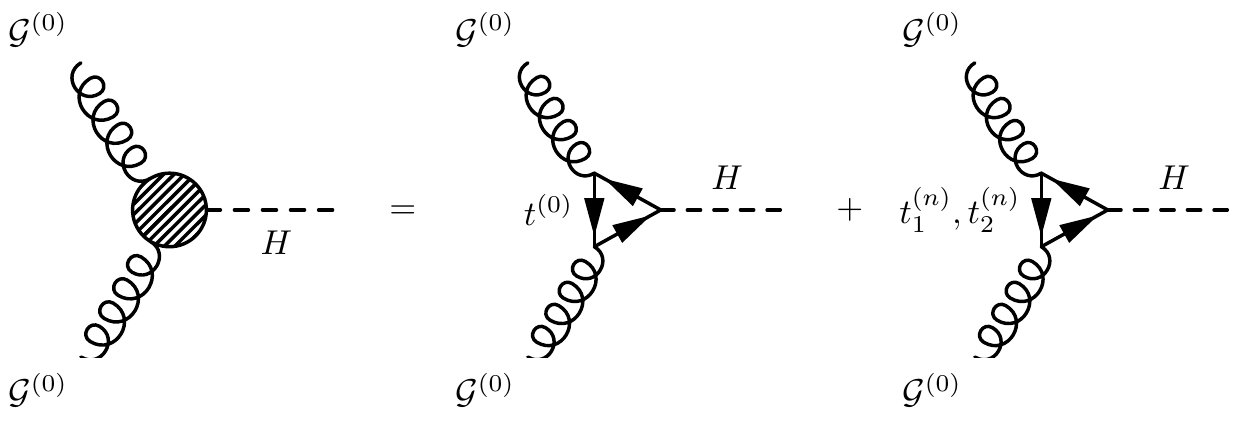}
\caption{The effective vertex which describes the Higgs production from the gluon fusion.}
\label{effectivevertex}
\end{figure}

For each model, we get {the following result, where $J_\text{model}$ indicates the corresponding loop function:}
\begin{align}
\hat\sigma^\text{model}_{gg\to H\to ZZ}
	&=	{\alpha_{{4s}}^2\over256\pi^3}
		\paren{m_Z\over v_\text{EW}}^4\,
		{1+\paren{\hat s-2m_Z^2}^2\over 8m_Z^4}\,
		{\hat s \over \paren{\hat s-M_H^2}^2+ \left({M_H\Gamma_H}\right )^2}
		\sqrt{1-{4m_Z^2\over \hat s}}\, 
		\cred{K}\ab{J_\text{model} \left ( \hat s \right )}^2,
	\label{sigma_model}
\end{align}
with
\begin{align}
 J_\text{mUED} (\hat s)
 	&= \skipped I\fn{ m_t^2 \over \hat s } +2 \sum_{n=1}^\infty \left( {m_t \over m_{t(n)}} \right )^2 I\fn{ {m_{t(n)}^2 \over \hat s} },\label{result_of_mUED}  \\
 J_\text{DH}(\hat s)
 	&=	\sqrt{2}\vep_1\sqrt{
			\left |
				\skipped  I\fn{ m_t^2 \over \hat s }
				+2\sum_{n=1}^\infty \left( {m_t \over m_{t(n)}} \right )^2 I\fn{ m_{t(n)}^2 \over \hat s }
				\right |^2 
 +\left |2 \sum_{n=1}^\infty \left ( {m_t \over m_{t(n)} } \right )^2 \tilde I\fn{m_{t(n)}^2 \over \hat s } \right |^2 }, \label{result_of_DH}
\end{align}
\begin{align} 
	{J_{T^2/Z_2}(\hat s) = J_{{RP}^2}(\hat s)}
		&= \skipped I\fn{ m_t^2 \over \hat s } +2 \sum_{{m+n \geq 1 \atop \text{or\ } m=-n \geq 1}}  \left( {m_t \over m_{t(m,n)}} \right )^2 I\fn{ m_{t(m,n)}^2 \over \hat s },
	\label{result_of_T2Z2}\\  
 {J_{T^2/Z_4}(\hat s)}
 	&= \skipped I\fn{ m_t^2 \over \hat s } +2 \sum_{m\geq1, n\geq0} \left( {m_t \over m_{t(m,n)}} \right )^2 I\fn{ m_{t({m,n})}^2 \over \hat s },
 \label{result_of_T2Z4}\\ 
	{J_{T^2/Z_2\times Z'_2}(\hat s)}
		&=	\skipped I\fn{ m_t^2 \over \hat s } +2 \sum_{m\geq0,n \geq 0, \atop (m,n) \not= (0,0)}  \left( {m_t \over m_{t(m,n)}} \right )^2 I\paren{ m_{t(m,n)}^2 \over \hat s },
	\label{result_of_T2Z2Z2}\\ 
 J_{S^2/Z_2}(\hat s)
 	&=	\skipped I\fn{ m_t^2 \over \hat s } +2 \sum_{j=1}^{j_\text{max}} \left( {m_t \over m_{t(j)}} \right )^2 n(j)\,I\paren{ m_{t(j)}^2 \over \hat s },
 \label{result_of_S2Z2}\\
 J_\text{PS}(\hat s) = {J_{S^2}(\hat s)}
 	&= \skipped I\fn{ m_t^2 \over \hat s } +2 \sum_{j=1}^{j_\text{max}} \left( {m_t \over m_{t(j)}} \right )^2 (2j+1)\,I\fn{ m_{t(j)}^2 \over \hat s },
 \label{result_of_PS}
\end{align}
where the KK top masses are given by
\begin{align}
 m_{t(n)}
 	&:= \sqrt{m_t^2 + {\frac{n^2}{R^2}}}
 	= 		\sqrt{m_t^2 + {{n^2}{M_\text{KK}^2}}}, \\
 m_{t(m,n)}
 	&:= \sqrt{m_t^2 + {\frac{m^2+n^2}{R^2}}}
 	=		\sqrt{m_t^2 + {{\paren{m^2+n^2} M_\text{KK}^2 }{}}}, \\
 m_{t(j)}
 	&:= \sqrt{m_t^2 + {\frac{j(j+1)}{R^2}}}
 	= 		\sqrt{m_t^2 + {\frac{j(j+1) M_\text{KK}^2}{2}}}.
\end{align}
Here the $M_\text{KK}$ is the first KK mass, which is written as
\al{
M_\text{KK}={1\over R}
}
for the compactifications based on $S^1/Z_2$, interval, and $T^2$ (namely, the mUED, DH, $T^2/Z_2$, $RP^2$, $T^2/Z_4$, and $T^2/(Z_2\times Z_2')$ models) and is written as
\al{
M_\text{KK}={\sqrt{2}\over R}
}
for the $S^2$-based ones (namely, the $S^2/Z_2$, PS {and $S^2$} models).
The gluon fusion process
for mUED in $S^1/Z_2$ is first shown in Ref.~\cite{Petriello:2002uu} and for $S^2/Z_2$ in Ref.~\cite{Maru:2009cu}.
Also it has been calculated for $T^2/Z_4$ and PS in Ref.~\cite{Nishiwaki:2011vi}.
The results for DH, $T^2/Z_2$, $RP^2$, $T^2/(Z_2\times Z_2')$ and $S^2$ are newly presented in this paper.
The factor $\sqrt{2} \vep_1$ in Eq.~(\ref{result_of_DH}) is equal to $2\sqrt{2}/\pi \sim 0.9$.
The origin of this suppression factor is non-orthonormality of mode functions on an interval.
In the case of $S^2$-based compactification, there are some degenerated states, the number of which is described with $n(j)$ on $S^2/Z_2$ and $(2j+1)$ on PS or {$S^2$} for each KK-index $j$.
The specific form of $n(j)$ for the orbifold $S^2/Z_2$ in Eq.~\eqref{result_of_S2Z2} is as follows:
\al{
n(j) = 
\begin{cases}
j+1 & \text{for} \ j : \text{even},  \\
j & \text{for} \ j : \text{odd}.
\end{cases}
}
Several comments are in order.
\begin{itemize}
\item The origin of the factor 2 in front of each KK summation is the fact there are both left and right handed (namely, vector-like) KK modes for each chiral quark zero mode.
\item All the KK contributions are positive and hence always enhance the Higgs production rate via the gluon fusion process, {except for the DH model in which the zero mode Higgs contribution is absent.}
\item {Each value of \skipped Yukawa couplings of KK quarks to the Higgs
is the same as that of the coupling between the corresponding zero mode fermion
and the Higgs. 
We only consider triangle loop diagrams
of the SM top quark and its KK excited modes
because their Yukawa coupling to the Higgs is dominant
compared to that of other fermions.}
\item In each KK summation infinite numbers of KK modes contribute to the process {in principle}. In 6D, these summations are divergent and a suitable scheme of regularization is required. $j_{\text{max}}$ in Eqs.~(\ref{result_of_S2Z2}) and (\ref{result_of_PS}) shows an upper bound of the summation over the index $j$.
{Further discussion will be shown in the following subsection.}
\item $K$ \cred{in Eq.~\eqref{sigma_model}} is the so-called K-factor, a phenomenological \cred{approximation} in order to naively take into account higher order QCD corrections. One may take \cred{$K\sim 2$ for Tevatron and $K\sim 1.5\text{--}1.6$ for LHC}, respectively~\cite{Djouadi:2005gi}.
\cred{
In the limit where the KK-loop is viewed as a contribution to the effective Higgs-gluon-gluon coupling, the QCD corrections to Higgs production are very similar between the SM and the new physics contributions. The reason is that the Higgs-gluon-gluon coupling always has the same structure, and only its coefficient changes. This is discussed in detail e.g.\ in Ref.~\cite{Djouadi:2005gj} in the context of SUSY (but it works the same way in UED as in SUSY). Therefore we have included a K factor also for the new physics terms as in Eq.~\eqref{sigma_model}.
}
\item 
{When the compactification radius is too large, namely, when the first KK $W$ is lighter than half the Higgs mass, 
the Higgs can decay into a pair of KK particles
and its decay width $\Gamma_H$ becomes broader.}
In this situation it {becomes harder} to find the evidence of the Higgs boson and {hence we restrict} ourselves within the region where the Higgs mass is \cred{smaller} than twice the first KK $W$ mass so that such a decay mode does not open up.
\item Although the one-loop gluon fusion process is the dominant production channel of the Higgs, its contribution to the Higgs total decay width is smaller at least by three orders of magnitude comparing to the decay into a $W$ pair in the case of the SM with $M_H \geq 300$\,GeV~\cite{Djouadi:2005gi}. Even after enhancement of $\mathcal{O}(10)$ from KK-top contributions, decay into gluon pair is still negligible.
\item In Eq.~\eqref{sigma_model}, the Higgs decay width is taken into account by the naive Breit-Wignar formula in the denominator. When the Higgs mass is large, say $M_H=700\GeV$, the Higgs decay width is as large as 180\GeV. In some literature, the expression $M_H\Gamma_H$ in the the Breit-Wignar formula is replaced by $\hat s\Gamma_H/M_H$. In Appendix~\ref{Two_point_function_and_width} we discuss reliability of our treatment. 
\end{itemize}

\subsection{UV cutoff scale in six dimensions}
\label{UV_cutoff_scale_in_six_dimensions}
In 6D UED models, since the gluon fusion {process} is 
UV divergent, we must consider upper limit of {the summations} of KK number in such {models}.\footnote{
In 5D UED model, we can execute this mode summation with no divergence.
}
First let us briefly review how the Naive Dimensional Analysis (NDA) is applied to the higher dimensional theory.
Following the {concept of} NDA, a loop expansion parameter $\epsilon$ in D-dimensional
SU(N) gauge theory at a scale $\mu$ is obtained as
\al{
\epsilon{(\mu)}
	&=	\frac{1}{2} \frac{2 \pi^{D/2}}{(2\pi)^D \Gamma(D/2)}
		N_g\,g_{Di}^{ 2}(\mu)\,\Lambda^{D-4},
			\label{expansion_parameter}
}
where $N_g$ is a group index, $g_{Di}$ is a {dimensionful} gauge coupling in $D$-dimensions and $\Lambda$ is a UV cutoff scale. The index $i$ is introduced to express the type of gauge interaction and the remaining part originates from $D$-dimensional momentum loop integral.
The cutoff scale $\Lambda$ is the scale where the perturbation breaks down $\epsilon(\Lambda)\sim 1$.

Precisely speaking, the dimensionful higher dimensional gauge coupling does not ``run.'' Let us explain what is meant by the running coupling in~\eqref{expansion_parameter}, basically following Ref.~\cite{Dienes:1998vg}.
When we consider a 6D theory $(D=6)$ with two {compact} spacial dimensions, an effective 4D gauge coupling $g_{4i}$ emerges after KK decomposition: $g_{4i} = {g_{6i}}/{\sqrt{V_2}}$, where $V_2$ is the volume of two extra dimensions.
Concretely, $V_2=(2\pi R)^2$ and $4\pi R^2$ for $T^2$ and $S^2$, respectively.
In this paper, we employ a bottom-up approach for the running gauge coupling. At energies below the first KK scale, theory is purely four dimensional (after integrating out all the massive modes of order KK scale) and the gauge coupling runs logarithmically. Let us then increase the energy scale. Every time we cross a KK mass scale, there open up the corresponding KK modes to run in the loops in the gauge boson two-point function. In all the scales, the theory is renormalizable and the running of the gauge coupling is logarithmic. However, due to the increase of the number of particles in the loops, the running of the gauge coupling becomes \emph{effectively} power-law at the energy scales much above the first KK scale. This way, we get the effective power-law running of the gauge coupling, within purely renormalizable approach. In this paper, we neglect possible threshold corrections at the UV cutoff scale, since we are interested in what is the highest possible $\Lambda$ that can be consistent with the low energy theory.

In the above stated strategy, we get the following running of the 4D effective gauge coupling strength $\alpha_{4i}(\mu)$
	\al{\cred{
	\alpha_{4i}^{\cred{-1}}(\mu) 
	\ 	\cred{\simeq}	\ \alpha_{4i}^{\cred{-1}}(m_Z) 
			-\frac{\textsf	{b}_i^{\text{SM}}}{2\pi}
				\ln{\mu \over m_Z}
			+2C\,{\textsf{b}_i^\text{6D}\over2\pi}
				\ln{\mu \over M_{\text{KK}}}
			-C\,\frac{\textsf{b}_i^{\text{6D}}}{2\pi}  
				\sqbr{ \left(  \frac{\mu}{M_\text{KK}} \right)^2 -1 },
			\label{6DcouplingRGequation}
	}}
where $C=\pi/2$ and 1 for $T^2$ and $S^2$, respectively.
{In Eq.~\eqref{6DcouplingRGequation}, we have approximated that all the masses are degenerate in each KK level. Depending on models, some fraction of the KK modes are projected out, but we assume that all of them contribute to the running, in order to give the most conservative upper bound on the UV cutoff scale.}
More detailed explanation is given in Appendix~\ref{Details_of_renormalizationgroupanalysis}.

\begin{table}
\begin{center}
\cred{
\begin{tabular}{|c||cc|cc|}
\hline
& \multicolumn{2}{|c|}{$T^2$-based}
& \multicolumn{2}{|c|}{$S^2$-based}\\
& max & min & max & min \\
\hline
KK index & $m^2+n^2 \leq  28$ & $m^2+n^2 \leq 10$ & $j(j+1) \leq 90$ & $j(j+1) \leq 30$\\
UV cutoff & 
	$\Lambda_{6D} \sim 5\MKK$ & 
	$\Lambda_{6D} \sim 3\MKK$ & 
	$\Lambda_{6D} \sim 7\MKK$ & 
	$\Lambda_{6D} \sim 4\MKK$ \\
\hline
\end{tabular}
\caption{Our choices of maximum and minimum upper bounds for KK indices and for the corresponding UV cutoff scale.}
\label{UEDcutoffvalues}
}
\end{center}
\end{table}

\cred{While the described procedure gives a reasonable estimate of the UV cutoff scale, one has to be aware that this is not much more than an order-of-magnitude estimate. We will plot our results for maximum and minimum values of the UV cutoff scale that are theoretically reasonable. In Table~\ref{UEDcutoffvalues}, we list our choice of bounds at which the KK mode summation is truncated to regularize the process.}

\subsection{Convolution of parton distribution}
Let us briefly review the standard prescription to estimate the event number of the Higgs production in $pp \to ZZ \to 4\ell$ (four leptons) via the gluon fusion process as a function of the invariant mass of $ZZ$, given the parton level cross section, where $4\ell$ denotes two pairs consisting of either $e^- e^+$ or $\mu^- \mu^+$, since a tau pair is less visible at the LHC.
That is, the final state is possibly $e^- e^+e^- e^+$, $\mu^- \mu^+\mu^- \mu^+$, or $e^- e^+\mu^- \mu^+$. 
Because of the large mass difference between Z boson and two electrons or
two muons, the subsequent processes: $Z\to e^+ e^-$ and $Z\to \mu^+ \mu^-$
are well treated with on-shell approximation.}

By using parton distribution function of gluon  $f_g(x, \hat s)$, 
we give the formula of the total cross section of $pp \to ZZ$:
\begin{equation}
 \sigma_{pp \to ZZ}^\text{model} (s)= \int_0^1 d\tau \hat\sigma^\text{model}_{gg\to H\to ZZ} (\tau s) L(\tau ,s),
\end{equation}
where we define
\begin{equation}
 L(\tau ,s)
 	:=	\int_{-\ln {1\over \sqrt{\tau}}}^{\ln {1\over \sqrt{\tau}}} dy f_g(\sqrt{\tau} e^y,\tau s) f_g(\sqrt{\tau} e^{-y},\tau s).
\end{equation}
The concrete form of $\hat\sigma_{pp \to ZZ}^\text{model}$ has been shown in Eqs.~\eqref{sigma_model}--\eqref{result_of_PS}.
%
The invariant mass of $ZZ$ is represented as $M^2_{ZZ} = \hat s = \tau s$.
Then, the differential cross section of $pp \to ZZ \to 4\ell$ as a function of $M_{ZZ}$ is written as
\begin{equation}
 {d\sigma_{pp \to ZZ \to 4\ell}^\text{model}(M_{ZZ}) \over dM_{ZZ} }
	=	{2M_{ZZ} \over s}
		\times \hat\sigma^\text{model}_{gg\to H\to ZZ}(M^2_{ZZ})
 		\times L\left ({M^2_{ZZ} \over s},s \right )
		\times 4 \Br(Z\to 2\ell)^2, 
\end{equation}
where $\Br(Z\to 2\ell):=\Br(Z\to e^+e^-)=\Br(Z\to \mu^+\mu^-)=0.034$ is the branching ratio.

\section{Numerical results}\label{numerical_results_section}
\begin{table}[t]
\begin{center}\begin{tabular}{|c||l|}\hline 
Geometry &  \multicolumn{1}{|c|}{Allowed region of  KK indices} \\
\hline 
$T^2/Z_2$ or $RP^2$& 
    $(m,n) = (1,0),\, (2,0),\, (3,0),\, (4,0),\, (5,0),$ \skipped \\
 & \phantom{$(m,n) =$} $(0,1),\, (1,1),\, (2,1),\, (3,1),\, (4,1),\, (5,1),$ \skipped \\
 & \phantom{$(m,n) =$} $(-1,2),\,  (0,2),\, (1,2),\, (2,2),\, (3,2),\, (4,2),$ \skipped \\
 & \phantom{$(m,n) =$} $(-2,3),\, (-1,3),\,  (0,3),\, (1,3),\, (2,3),\, (3,3),\, (4,3),$ \skipped \\ 
 & \phantom{$(m,n) =$} $(-3,4),\, (-2,4),\, (-1,4),\,  (0,4),\, (1,4),\, (2,4),\, (3,4),$ \skipped \\ 
 & \phantom{$(m,n) =$} $\skipped (-1,5),\,  (0,5),\, (1,5),$ \skipped \\ 
 & \phantom{$(m,n) =$} $(1,-1),\, (2,-1),\, (3,-1),\, (4,-1),\, (5,-1),$ \skipped \\ 
 & \phantom{$(m,n) =$} $(2,-2),\, (3,-2),\, (4,-2),$ \skipped \\
 & \phantom{$(m,n) =$} $(3,-3),\, (4,-3),$ \skipped \\
\hline
$T^2/(Z_2 \times Z'_2)$ & 
    $(m,n) = (1,0),\, (2,0),\, (3,0),\, (4,0),\, (5,0),$ \skipped \\
 & \phantom{$(m,n) =$} $(0,1),\, (1,1),\, (2,1),\, (3,1),\, (4,1),\, (5,1),$ \skipped \\
 & \phantom{$(m,n) =$} $(0,2),\, (1,2),\, (2,2),\, (3,2),\, (4,2),$ \skipped \\
 & \phantom{$(m,n) =$} $(0,3),\, (1,3),\, (2,3),\, (3,3),\, (4,3),$ \skipped \\ 
 & \phantom{$(m,n) =$} $(0,4),\, (1,4),\, (2,4),\, (3,4),$ \skipped \\ 
 & \phantom{$(m,n) =$} $(0,5),\, (1,5),$ \skipped \\ 
\hline
$T^2/Z_4$ & 
    $(m,n) = (1,0),\, (2,0),\, (3,0),\, (4,0),\, (5,0),$ \skipped \\
 & \phantom{$(m,n) =$} $\skipped (1,1),\, (2,1),\, (3,1),\, (4,1),\, (5,1),$ \skipped \\
 & \phantom{$(m,n) =$} $(1,2),\, (2,2),\, (3,2),\, (4,2),$ \skipped \\
 & \phantom{$(m,n) =$} $(1,3),\, (2,3),\, (3,3),\, (4,3),$ \skipped \\ 
 & \phantom{$(m,n) =$} $(1,4),\, (2,4),\, (3,4),$ \skipped \\ 
 & \phantom{$(m,n) =$} $(1,5),$ \skipped \\ 
\hline
$S^2/Z_2$ & \hspace{6mm} $j= 1 \sim \cred{9}$ ($n(j) = j+1$ for $j$:\,even or $n(j) = j$ for $j$:\,odd)\\
\hline
PS or {$S^2$} & \hspace{6mm} $j= 1 \sim \cred{9}$ ($2j+1$ degenerated states)\\
\hline
\end{tabular}
\caption{{The region of KK summation \cred{for the maximum UV cutoff}.}}
\label{KKsumtable}
\end{center}\end{table}

We use the top quark mass $m_t =172$\,GeV and the LO QCD running coupling.
We apply the CTEQ5 LO PDF of gluon~\cite{Lai:1999wy}. 
In our analysis, we consider only KK-scales large enough not to open up the Higgs decay channel into a pair of KK-top, $Z$, and $W$ particles.
When the decay channel into two KK particles opens up, the Higgs resonance tends to become too broad and hard to be seen at Tevatron and LHC. 
\cred{We list in Table~\ref{KKsumtable} the KK modes that satisfy the maximum cutoff criterion given in Table~\ref{UEDcutoffvalues}.}

\subsection{Tevatron}
\begin{figure}[t]
\centering
\includegraphics[width=40em, bb= 0 0 953 380, clip]{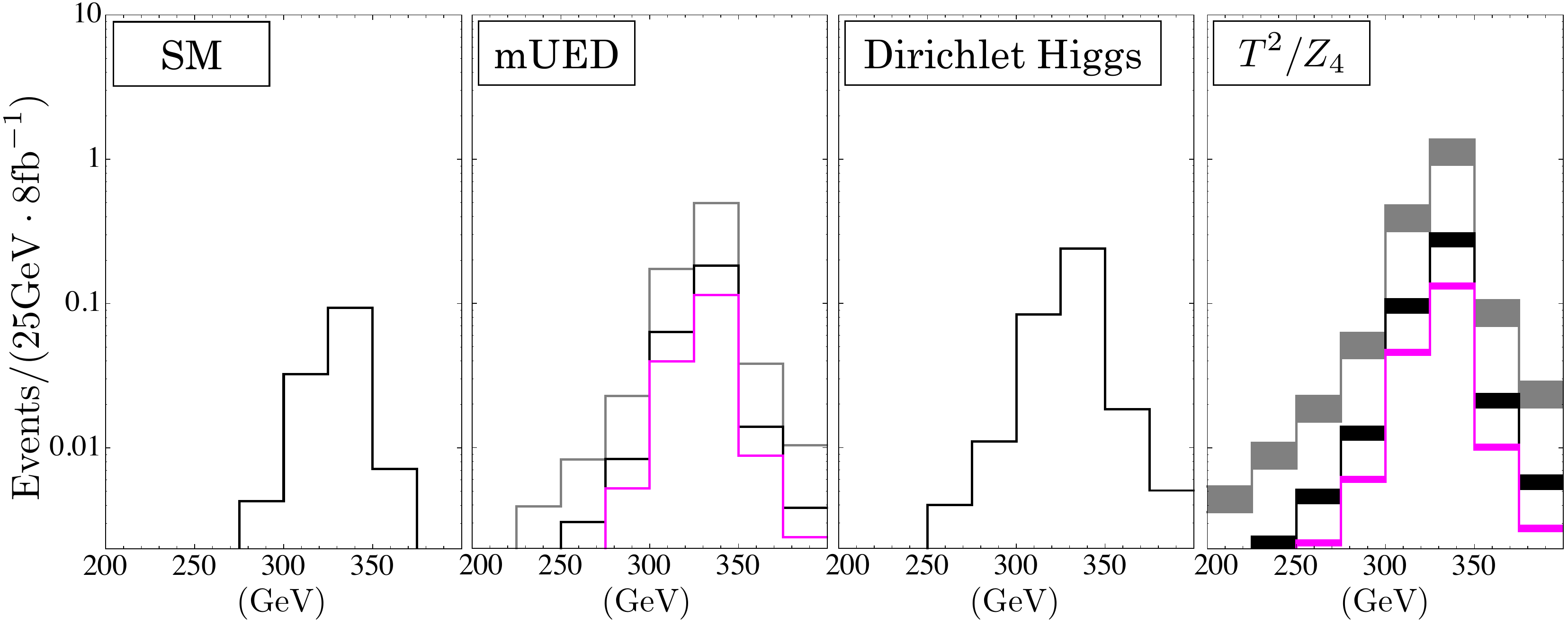}
\includegraphics[width=40em, bb= 0 0 952 379, clip]{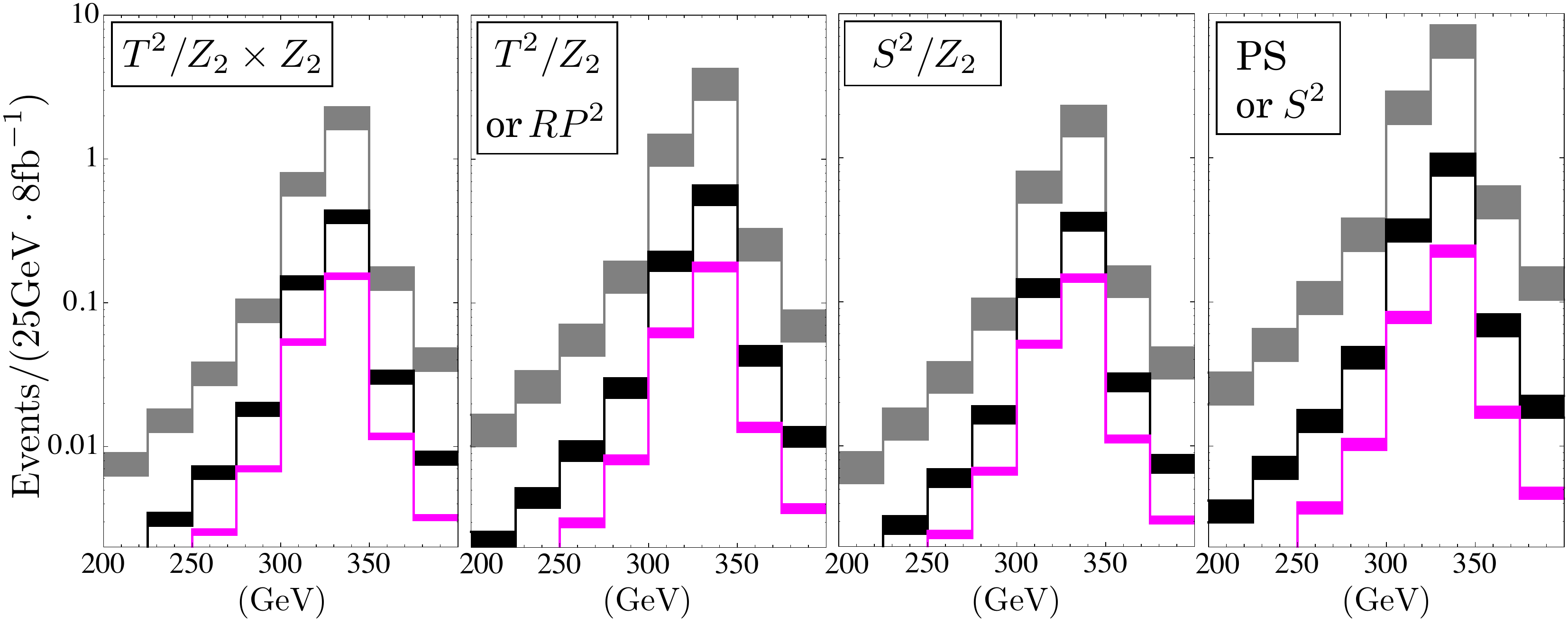}
\caption{
The $H\to ZZ\to4\ell$ event number per each \cred{25}\,GeV bin of $M_{ZZ}$ for $M_H=330$\,GeV expected at Tevatron with an integrated luminosity $8\fbinv$ at $\sqrt{s}=1.96$ TeV. The \cred{grey, black, and magenta} lines represent the expected event number with $M_\text{KK} = 200,\,400$, and $\cred{800}$\,GeV, respectively.
\cred{For 6D UED models, we consider dependency on the UV cutoff, whose range is from minimum (lower side of band) to maximum (upper sider of band) given in Table~\ref{UEDcutoffvalues}.}}
\label{TevaRes}
\end{figure}

We evaluate the $H\to ZZ\to4\ell$ event number per each \cred{25}\,GeV bin of $M_{ZZ}$, expected at Tevatron with an integrated luminosity $8\fbinv$ at $\sqrt{s}=1.96$ TeV in the above mentioned UED models. 
We show the results for the Higgs mass $M_H=330$\,GeV.
We consider the KK scales $M_\text{KK} = 200,\,400$, and $\cred{800}$\,GeV, except for the Dirichlet Higgs (DH) model which does not have a zero-mode Higgs. 
For the DH model, we take first-KK Higgs as the Higgs field, that is, $M_\text{KK} =M_H=330$\,GeV.\footnote{
Note that the coupling of the Dirichlet Higgs to the SM modes is decreased by a factor 0.9 compared to the SM Higgs. 
}
The results are shown in Fig.~\ref{TevaRes}.
We see that the event number is enhanced in all the UED models from the SM one. 
In particular, the 6D Projective Sphere model can give a large enhancement in the Higgs production by a factor as large as hundred \cred{compared to the SM} when the KK-scale is low at \cred{200}\,GeV.\footnote{
Even if the one-loop $H\to gg$ decay rate is enhanced by the factor 100 from the SM one, it is still subdominant compared with the tree-level $H\to WW$~\cite{Djouadi:2005gi} and we neglect its contribution to the total decay width of the Higgs.
}
Two $ZZ\to4\ell$ events are observed in a 300--350\,GeV bin at CDF with the integrated luminosity $4.8\fbinv$~\cite{ZZ4l} and two such events are observed in 325--375\,GeV bins at D0 with $6.4\fbinv$~\cite{Abazov:2011td}.
Recently two more events are reported around 330\,GeV~\cite{CDF_latest}.
However, the above large cross section to explain the Tevatron data is found to be inconsistent with the LHC data~\cite{ATLAS_latest,CMS_latest}.


\begin{figure}[t]
\centering
\includegraphics[width=40em, bb= 0 0 953 381, clip]{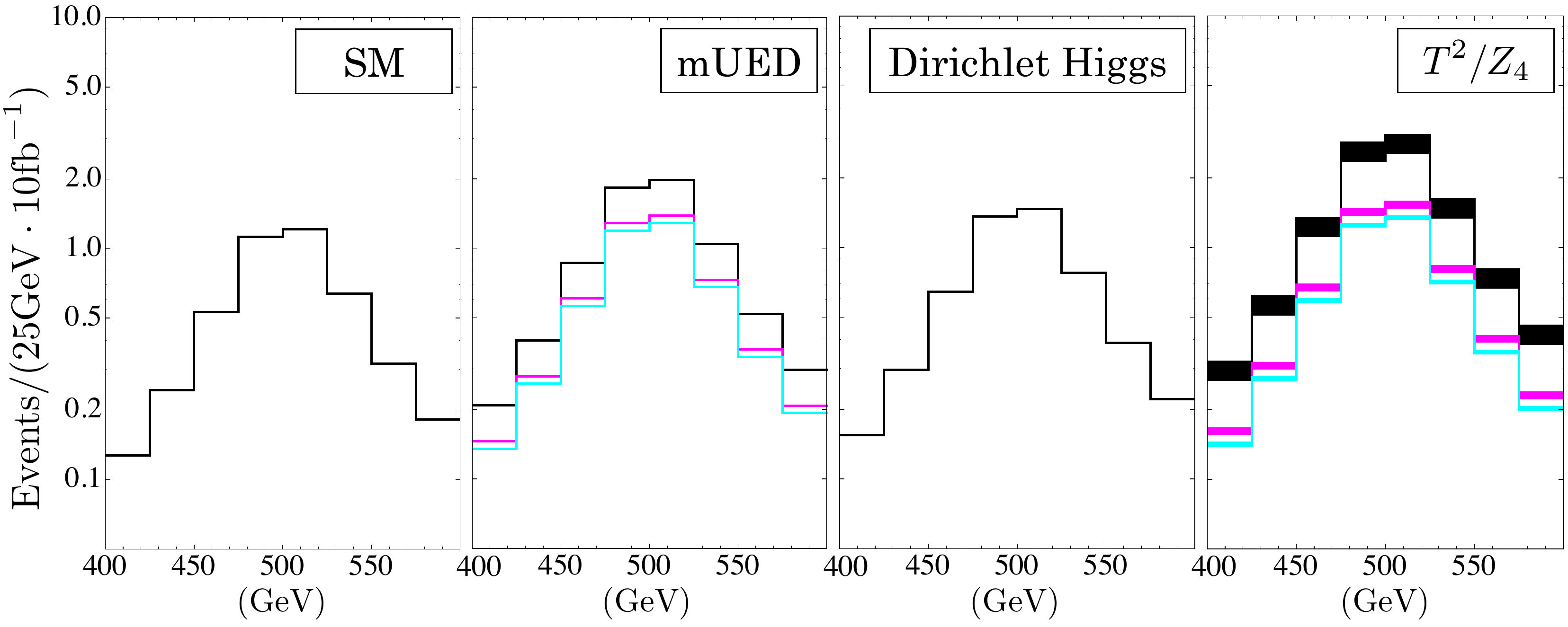}
\includegraphics[width=40em, bb= 0 0 952 380, clip]{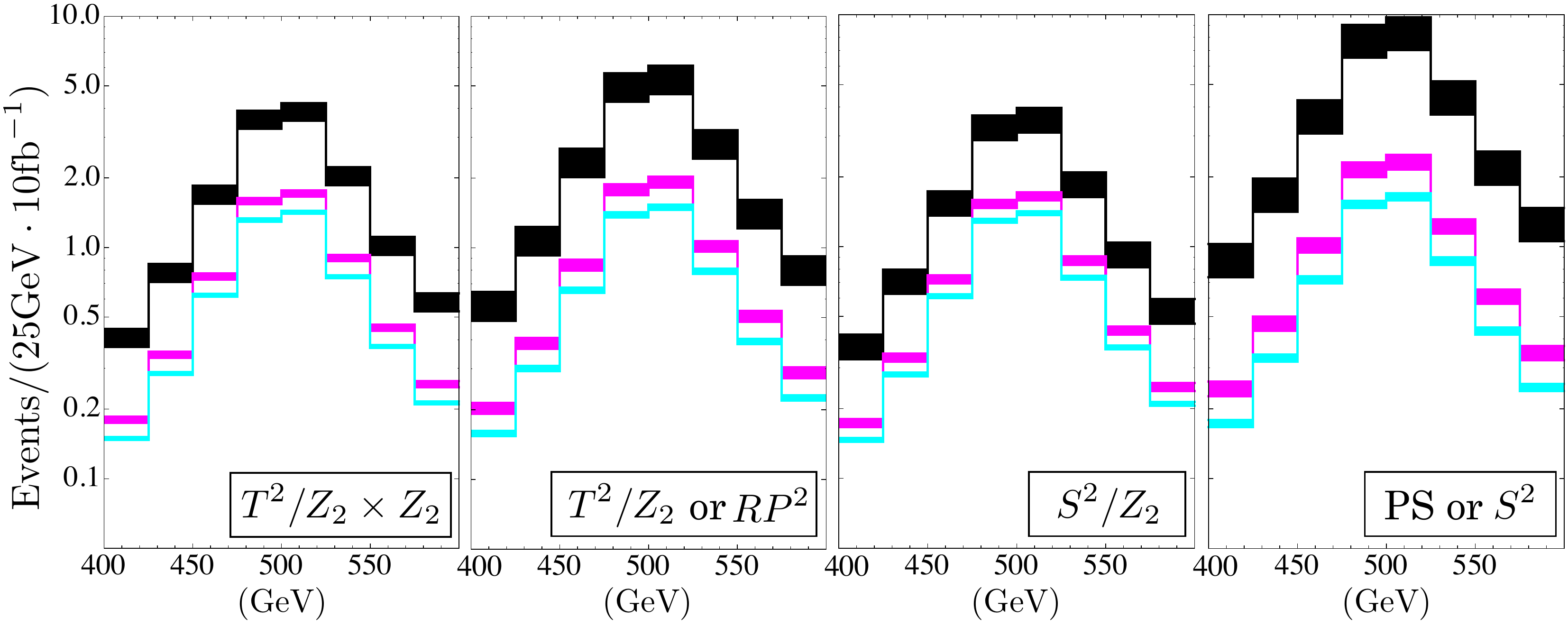}
\caption{The $H\to ZZ\to4\ell$ event number per each 25\,GeV bin of $M_{ZZ}$ for $M_H=500$\,GeV, expected at LHC with an integrated luminosity $\cred{10}\fbinv$ at $\sqrt{s}=7$ TeV. The \cred{black, magenta, and cyan} lines represent the expected event number with \cred{$M_\text{KK} = 400,\,800$, and $1200$}\,GeV, respectively.
\cred{Dependence on the 6D UV cutoff scale is shown the same as in Fig.~\ref{TevaRes}.}
}
\label{LHCRes500}
\end{figure}

\begin{figure}[t]
\centering
\includegraphics[width=40em, bb= 0 0 953 381, clip]{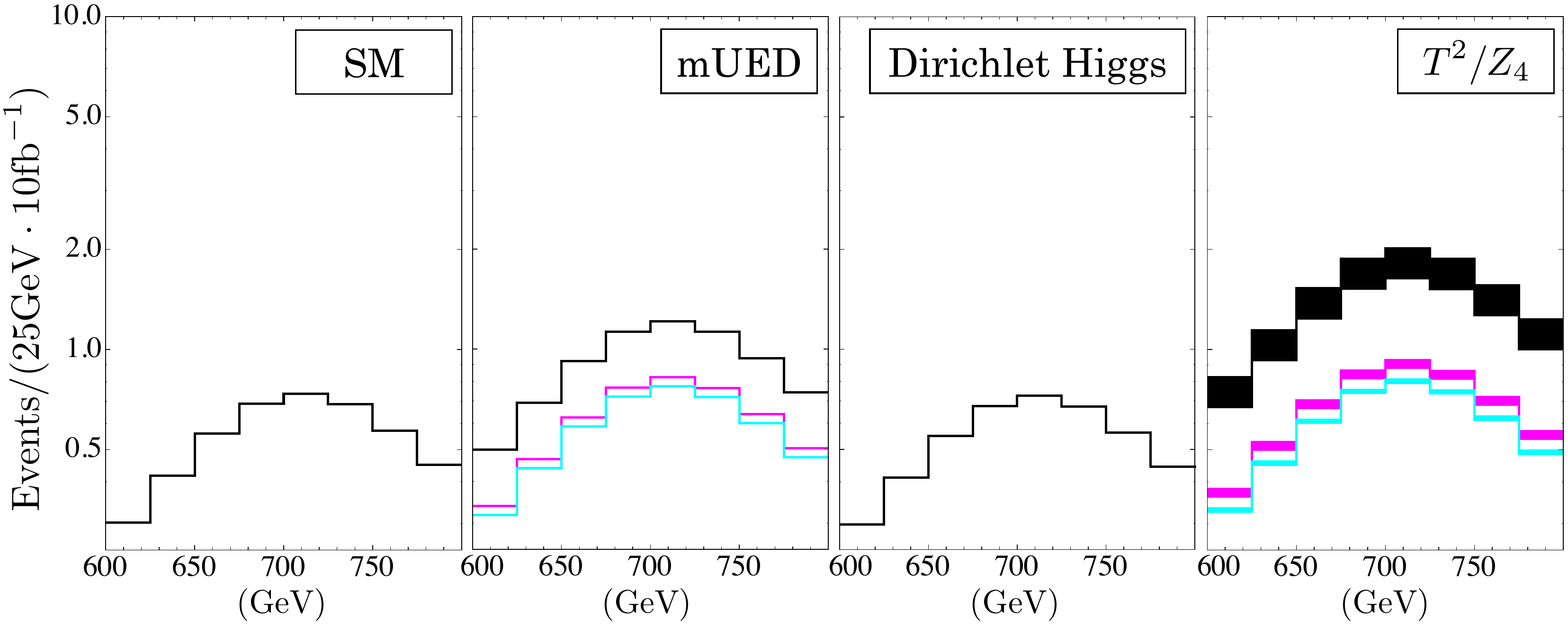}
\includegraphics[width=40em, bb= 0 0 952 380, clip]{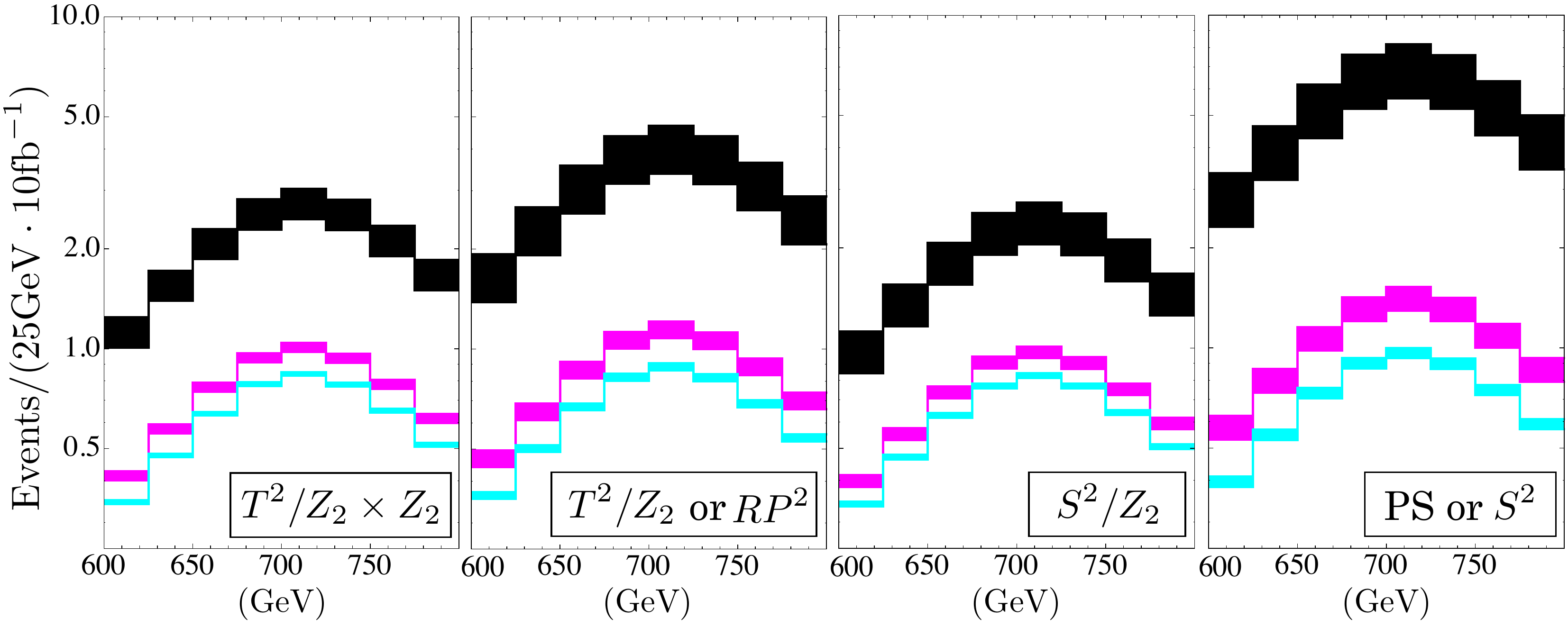}
\caption{The $H\to ZZ\to4\ell$ event number per each 25\,GeV bin of $M_{ZZ}$ for $M_H=700$\,GeV, expected at LHC with an integrated luminosity $\cred{10}\fbinv$ at $\sqrt{s}=14$ TeV. The \cred{black, magenta, and cyan} lines represent the expected event number with \cred{$M_\text{KK} = 400,\,800$, and $1200$}\,GeV, respectively.
\cred{Dependence on the 6D UV cutoff scale is shown the same as in Fig.~\ref{TevaRes}.}
}
\label{LHCRes700}
\end{figure}

\subsection{\cred{LHC}}

We plot the event number of $H\to ZZ\to4\ell$ for the Higgs mass \cred{$M_H=500\GeV$} at the LHC with an integrated luminosity $\cred{10}\fbinv$ at $\sqrt{s}=7$ TeV in \cred{Fig.~\ref{LHCRes500}. } When $\sqrt{s}=7\TeV$, we have checked that we cannot see sizable number of events for all the UED models with $M_H=700\GeV$ even for an integrated luminosity $10\fbinv$, expected by the end of 2012 after which the LHC is planned to be (shut-down for a year and then) upgraded to 13--14\,TeV.
Therefore, we show corresponding results for $M_H=700\GeV$ at $\sqrt{s}=14\TeV$ with an integrated luminosity $\cred{10}\fbinv$ in Fig.~\ref{LHCRes700}.
We have chosen several KK scales:
\skipped 400, \skipped 800, \cred{1200}\,GeV. \skipped
We show our plots in logarithmic scale so that one can easily see the results for different luminosities by simply shifting them upward/downward.

\skipped

In Fig.~\ref{LHCRes500} with an integrated luminosity $\cred{10}\fbinv$,     \skipped
we can see in total a few events in 5D UED models (mUED and DH) and $\mc O(1)$--$\mc O(10)$ events per each 25\,GeV bin for 6D UED models.
Note that even if we can only see at best in total a few events for 5D UED by the end of 2012, this $ZZ\to4\ell$ channel is virtually background free at 500\,GeV and the result would be still significant.

In Fig.~\ref{LHCRes700} for $M_H=700\GeV$, we have plotted the results for the upgraded energy $\sqrt{s}=14\TeV$.
We see that even with the integrated luminosity $10\fbinv$, 6D UED models can have a few events per each 25\,GeV bin if the KK scale is relatively low $M_\text{KK}=400\GeV$. \skipped The 6D $T^2/Z_2$, $RP^2$, $S^2$ and PS models can have \cred{in total} few events \skipped for higher KK mass $M_\text{KK}=\cred{800}\GeV$.
When the integrated luminosity adds up to $100\fbinv$, we can see a few events per each bin for 5D mUED and DH models, and even for the SM (though the SM itself cannot satisfy the electroweak constraints on the $S,T$ parameters, contrary to the UED models).
We see that the Dirichlet Higgs model has slightly smaller cross section than the Standard Model. This is because the KK scale is fixed to be large, 700\,GeV, and hence the enhancement of the KK top loop is small, while the Yukawa coupling of the (first KK) Higgs to the top quark is decreased by the factor $2\sqrt{2}/\pi\simeq 0.9$.

Let us emphasize that the enhancement of Higgs production $gg\to H$ in UED does not depend on the details of the model such as the mass structure at the orbifold fixed points.
Parameter dependence is only on the Higgs mass and the KK scale.
In this sense, this Higgs channel signal is complementary to the direct search of the KK modes decaying into LKP~\cred{\cite{Murayama:2011hj,Tobioka:master}}, which is nice because of the directness but is dependent on the details of KK mass splitting from the boundary terms.



\section{Summary and discussions}
We have presented a review on the known 5D and 6D UED models focusing on the relevant part to the gluon fusion process.
We have explained our computation of the gluon fusion process including the KK top-quark loops, which is new for the $T^2/Z_2$, $T^2/(Z_2\times Z_2')$, $RP^2$, and $S^2$ UED models.
For 6D UED models, we have shown an NDA analysis of the highest possible UV cutoff scale in the $S^2$-based compactification, extending the analysis of Ref.~\cite{Dienes:1998vg,Bhattacharyya:2006ym} on the $T^2$ compactification.

\skipped
For \skipped Higgs mass $M_H=500\GeV$, we can see a few (virtually background free) $H\to ZZ\to 4\ell$ events in 5D UED models with $10\fbinv$ of integrated luminosity. The 6D UED models can further exhibit the shape of the resonance if the KK scale is relatively low.
When Higgs mass is as large as $M_H=700\GeV$, we found no parameter region that can be seen within the integrated luminosity of $\mc O(10)\fbinv$ at $\sqrt{s}=7\TeV$. We have also studied the event rate for the upgraded energy $\sqrt{s}=14\TeV$ when $M_H=700\GeV$. We see that the 5D and 6D UED models typically require $100\fbinv$ and $10\fbinv$ of data, respectively, in order to establish the existence of the resonance.
As is reported in Ref.~\cite{Nishiwaki:2011vi}, the 6D UED model on the Projective Sphere (or $S^2$) shows the greatest enhancement of the Higgs production via gluon fusion process, among all the known UED models.

Let us again emphasize that 
the presented Higgs signal of UED
needs only the Higgs mass and KK scale as input parameters,
is independent of the detailed KK mass splitting, and hence
is unaffected by the boundary/fixed-point mass structures. Therefore, this Higgs signal of UED is complementary to the direct KK resonance production and to the dark matter signal.
%
As the enhancement of the Higgs production via the gluon fusion process can be so large, recent data from the LHC~\cite{ATLAS_latest,CMS_latest} can already significantly exclude the parameter space of the UED models. This analysis \cred{is} presented in a separate publication~\cite{NOOW_letter}.
\cred{We will also show a combined bound from the triviality  and the electroweak precision constraints in addition to that of Ref.~\cite{NOOW_letter}. (The triviality bound would lower the maximum allowed UV cutoff scale when the Higgs mass is heavy.)}
%
In this paper, we have studied the cleanest possible signature $H\to ZZ\to 4\ell$. It is expected that a combined analysis including other decay channels such as $H\to WW$ and $H\to ZZ\to \ell\ell\nu\nu$ will provide a large gain over all individual analyses~\cite{CMS_latest}. Such a combined analysis for the UED models will be presented elsewhere.

\subsection*{Acknowledgment}
We are grateful to Shoji Asai for valuable comments and thank
Abdelhak Djouadi,
Kaoru Hagiwara,
Kazunori Hanagaki,
Shigeki Matsumoto,
Kazuki Sakurai, and
\cred{Kohsaku} Tobioka for helpful communications.
\cmagenta{We appreciate Referee's thorough reading and constructive comments.}
The work of K.O.\ is partially supported by Scientific Grant by Ministry of Education and Science (Japan), Nos.~20244028, \cred{23104009} and 23740192.

\appendix
\section*{Appendix}

\section{Feynman rules for Dirichlet Higgs model
\label{DH_calculation_in_appendix}}
In this appendix, we describe the Feynman rules that are necessary for our computation.
%
%
%
The mass terms of KK fermions are 
\al{
- \sum_{n=1}^{\infty}
\bb \ol{Q_{t}} & \ol{U_t} \eb^{(n)}
\bb \frac{n}{R} & m_t \\
m_t & -\frac{n}{R} \eb
\bb {Q_t} \\ {U_t} \eb^{(n)}, 
}
where $Q_t$ is an upper component of the quark doublet in third generation
and $U_t$ is the top quark singlet.
Transforming each KK states by the following unitary transformation including chiral rotation:
\al{
\bb
t_1\\
t_2
\eb^{(n)}
	&=	\bb
		\gamma^5\\
			&	1
		\eb
		\bb
		\cos\alpha^{(n)}	&	-\sin\alpha^{(n)}\\
		\sin\alpha^{(n)}	&	\cos\alpha^{(n)}
		\eb
		\bb
		Q_t\\
		U_t
		\eb^{(n)},
}
we can obtain the ordinary diagonalized Dirac mass terms, where $t_1^{(n)}$ and $t_2^{(n)}$ are mass eigenstates of n-th KK top quarks and each mixing angle $\alpha^{(n)}$ is determined to be $\cos{2 \alpha^{(n)}} = (n/R)/\sqrt{m_t^2 + n^2/R^2}$,\ $\sin{2 \alpha^{(n)}} = m_t/\sqrt{m_t^2 + n^2/R^2}$.
Each KK state is twofold degenerate and n-th KK top mass is $m_{t,(n)} = \sqrt{m_t^2 + n^2/R^2}$.
The corresponding interaction terms are 
\al{
\hspace{-16mm}
\mc L_\text{KK top}
	&=	- i g_{{4s}}
		\sum_{n=1}^{\infty}
		\bb \ol{t_1} & \ol{t_2} \eb^{(n)}
		\gamma^{\mu} \G_{\mu}^{(0)}
		\bb t_1 \\ t_2 \eb^{(n)}
		\nn
	&\quad
		- {m_t\over v_\text{EW}}
		\sum_{n,m=1}^{\infty}
		\bb \ol{t_1} & \ol{t_2} \eb^{(n)}
		H^{(m)}
		\left( \sqrt{2} \vep_m + \frac{1}{\sqrt{2}} (\vep_{2n+m} - \vep_{2n-m}) 		\gamma^5 \right)
		\bb \sin{2 \alpha^{(n)}} & -\gamma^5\cos{2 \alpha^{(n)}} \\
		\gamma^5\cos{2 \alpha^{(n)}}  & \sin{2 \alpha^{(n)}} \eb
		\bb t_1 \\ t_2 \eb^{(n)},
}
where $g_{{4s}}$ is a dimensionless 4D $SU(3)_C$ coupling constant 
and $v_{\text{EW}}$ is \cred{the} 4D Higgs vacuum expectation value
which appears
after KK expansion.
$\G^{(0)}_{\mu}$ is massless gluon and $H^{(m)}$ is $m$-th KK Higgs bosons.
{The concrete shape of the factor of $\sqrt{2} \vep_n$ is $2\sqrt{2}/n\pi$, whose origin is
the non-orthonormality of mode functions in Dirichlet Higgs model.}
In Dirichlet Higgs model there is no zero mode Higgs
because of choosing Dirichlet boundary condition in Higgs field.
The first KK Higgs boson behaves like a heavy SM Higgs except that its interaction with the SM fields are multiplied by $\sqrt{2}\vep_1$. 
The explicit form of Feynman rules is:

\al{
\raisebox{-8mm}{\includegraphics[width=8em, bb= 0 0 64 24, clip]{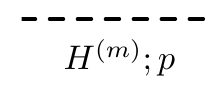}}
	&\quad=	{i\over -p^2 - \paren{m / R}^2+i\epsilon},\\
\raisebox{-8mm}{\includegraphics[width=7.5em, bb= 0 0 61 27, clip]{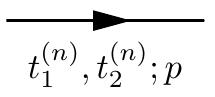}}
	&\quad=	{i\over  i\slashed{p}  +\sqrt{m_t^2 + n^2/R^2} +i\epsilon},\\
\raisebox{-7mm}{\includegraphics[width=20em, bb= 0 0 246 65, clip]{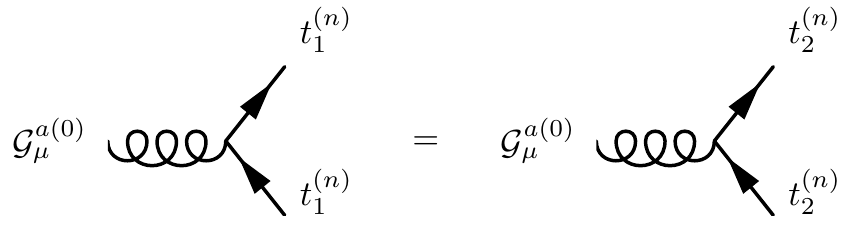}}
	&\quad=	{ g_{4s} \gamma_{\mu} \sqbr{ \frac{\lambda^a}{2} } },\\	
\raisebox{-12mm}{\includegraphics[width=20em, bb= 0 0 243 68, clip]{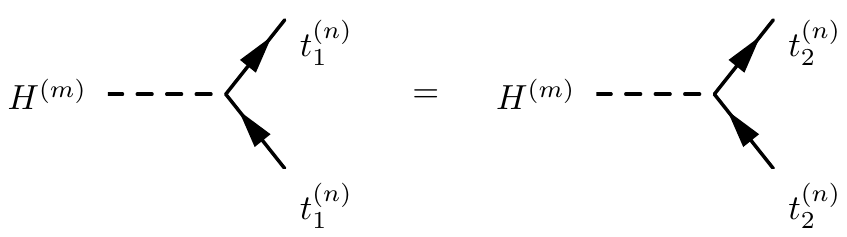}}
	&\quad=	{-i \frac{m_t}{v_{\text{EW}}} \left( \sqrt{2} \vep_m + \frac{1}{\sqrt{2}} (\vep_{2n+m} - \vep_{2n-m}) \gamma^5 \right) \sin{2\alpha^{(n)}}},
}

\noindent
where $a$ is a gluon (adjoint) color index and $\lambda^a$ are the Gell-Mann matrices. It is noted that we can find an interaction which is proportional to $\gamma^5$ matrix at the KK Higgs Yukawa couplings to KK top quarks, which generates another type of contribution to the Higgs production through the gluon fusion process.\footnote{
This situation is similar to the famous fact that $\pi^0 \rightarrow 2\gamma$ decay is enhanced through chiral anomaly~\cite{Steinberger:1949wx}.
}
This type of contribution do not exist in ordinary 5D or 6D UED models.

\section{UV cutoff based on RGE analysis in KK picture}
\label{Details_of_renormalizationgroupanalysis}
In this section, we show the details of the renormalization group analysis, based on the strategy stated in Section~\ref{UV_cutoff_scale_in_six_dimensions} that follows Ref.~\cite{Dienes:1998vg}.
We note that we do not need any regularization of the infinite KK sum, employed in Ref.~\cite{Dienes:1998vg}, in our bottom-up approach.
%
A higher-dimensional gauge theory is equivalent to the corresponding 4D theory with infinite tower of KK modes.
It suggest that all we have to do for deriving running effect of the 4D effective gauge coupling at leading order is to count the d.o.f.\ of fields whose masses are lower compared to a reference energy $\mu$.
This important information is encoded into the coefficient $\textsf{b}_i$.
The formula for the running coupling of 4D $SU(N)$ or $U(1)$ gauge theory is well-known
as follows:
\al{
\frac{d}{d \ln{\mu}} \alpha_{4i}^{-1} &= -\frac{1}{2\pi} \textsf{b}_i ,
\label{4Dcouplingequation}
}
\al{
\textsf{b}_{SU(N)} &=  \paren{ -\frac{11}{3}\cmagenta{\sum_{\text{4D vectors}}}C_2(\text{Adjoint}) + \frac{2}{3} \sum_{\text{4D Weyl} \atop \text{fermions}} C\paren{r} + \frac{1}{3} \sum_{\text{4D Higgs}} C\paren{r} + \frac{1}{6} \sum_{\text{4D adjoint} \atop \text{scalars}} C_{\cmagenta 2}(\text{Adjoint})  }, 
	\label{bSUN}\\
\textsf{b}_{U(1)} &= \paren{
\frac{2}{3} \sum_{\text{4D Weyl} \atop \text{fermions}} Y_f^2
+ \frac{1}{3} \sum_{\text{4D Higgs}} Y_H^2},      }
where $\alpha_{4i}$ is a 4D gauge coupling strength, whose group is discerned by $i$;
$C\paren{r}$ is defined by $\tr\sqbr{T^a_r T^b_r} = C\paren{r} \delta^{ab}$
for each representation of $SU(N)$ group 
and the specific value is $1/2$ \skippedMagenta for the fundamental \skippedMagenta representation;
$C_2(\text{Adjoint})$ is the quadratic Casimir operator for the adjoint representation,
whose value is $N$; $Y_f\, (Y_H)$ shows $U(1)$ charge of 4D Weyl fermion\,(Higgs).
\cred{The contribution from KK vector bosons is included in the first term in Eq.~\eqref{bSUN}.}
We note that a 6D gauge field has two extra dimensional components 
and that a 4D adjoint scalar is left as a physical mode after the KK decomposition.
\cred{(\skippedMagenta When counting the number of adjoint scalar degrees of freedom in the running~\eqref{4Dcouplingequation}, it is two rather than one \cmagenta{in our treatment}.)}

We can obtain the solution of Eq.~(\ref{4Dcouplingequation}) by the integration of the both sides over the region $[m_Z,\mu]$.
The essential point is that the coefficient $\textsf{b}_i$ becomes altered when the reference energy $\mu$ cross a threshold and the number of the effectively massless d.o.f.\ of fields under the $\mu$ changes.
Neglecting the KK mass splitting from the electroweak symmetry breaking, the threshold correction from $l$-th KK particles arrises simultaneously when $\mu$ exceeds the value of $l$-th KK mass.
This approximation simplifies the calculation 
to estimate the effect from $l$-th KK particles all at once.
The values of $\textsf{b}_i$ in the zero modes\,($\textsf{b}_i^{\text{SM}}$) and $l$-th KK modes\,($\textsf{b}_i^{\text{6D}}$)
from the bulk SM matter contents are summarized in Table~\ref{RGequation_coefficients}.\footnote{
Note that we do not employ the GUT normalization for the $U(1)_Y$ coupling and the beta function.
}
\begin{table}[t]
\begin{center}
\begin{tabular}{|c|c|c|}
\hline 
gauge group & SM contribution ($\textsf{b}^{\text{SM}}_i$) & KK contribution ($\textsf{b}^{\text{6D}}_i$) \\
\hline 
$SU(3)_C$ &  $\displaystyle -7$ & $-2$ \\
$SU(2)_W$ & $\displaystyle -{19}/{6}$ & $\displaystyle {3}/{2}$ \\
$U(1)_Y$ & $\displaystyle {41/6}$ & $\displaystyle {27}/{2}$ \\
\hline
\end{tabular}
\caption{\cred{RGE coefficients in Eq.~(\ref{6DcouplingRGequation}).}}
\label{RGequation_coefficients}
\end{center}
\end{table}%

This way, the shape of $\alpha_{4i}(\mu)$ is determined as
\al{
\alpha_{4i}^{\cred{-1}}(\mu) = \alpha_{4i}^{\cred{-1}}(m_Z) - \frac{\textsf{b}_i^{\text{SM}}}{2\pi} \ln{\mu \over m_Z}
- \frac{\textsf{b}_i^{\text{6D}}}{2\pi} \sum_{l} \ln\frac{\mu}{m_{(l)}},
}
where $m_{(l)}$ shows $l$-th KK mass and the upper bound of the summation
is
\al{
m_{(l)} \leq \mu.
}
In the case of $S^1$ in five dimensions, where the spectrum of KK masses is equally-spaced, the above calculation is executed with no difficulty. 
In 6D cases, by contrast, the KK mass spectrum is not equally-spaced and  we use the following approximation.


\begin{description}
\item[$\square$ $T^2$-case:]
\noindent
In the $T^2$-case, the form of $(m,n)$-th KK mass is $m_{(m,n)} = \sqrt{m^2+n^2}/R$ and
eventually that of the first\,(lightest) KK mass is $M_\text{KK} = 1/R$.
The exact form of the summation is as follows:
\al{
\sum_{(m,n)} \ln{\mu \over \sqrt{m^2+n^2} M_\text{KK}} & &\text{for} \quad 
1 \leq m^2+n^2 \leq \paren{\mu \over M_\text{KK}}^2. 
}
We approximate the summation over $m$ and $n$ by integral over $r$ and $\theta$ in two-dimensional polar coordinates:
\al{
\sum_{(m,n)} \paren{\ln{\mu \over M_\text{KK}} -\frac{1}{2} \ln\paren{m^2+n^2} }
&\simeq
\int_1^{\mu/M_\text{KK}} 2\pi dr \cdot r \paren{ \ln{\mu \over M_\text{KK}} - \ln{r}} \notag \\
&= \frac{\pi}{2} \sqbr{ \left(  \frac{\mu}{M_\text{KK}} \right)^2 -1 \cred{-2\ln{\mu\over M_\text{KK}}}},
}
which results in
\al{
\alpha_{4i}^{\cred{-1}}(\mu) \ \cred{\simeq}\  \alpha_{4i}^{\cred{-1}}(m_Z) - \frac{\textsf{b}_i^{\text{SM}}}{2\pi} \ln{\mu \over m_Z}
- \frac{\textsf{b}_i^{\text{6D}}}{2\pi} \cdot 
\frac{\pi}{2}
	\sqbr{
		\paren{\frac{\mu}{M_\text{KK}}}^2	-1
		\cred{	{}
			-2\ln{\mu\over M_\text{KK}}	}
		}.
	\label{RGE_for_T2}
} 
\item[$\square$ $S^2$-case:]
\noindent
In the $S^2$-case, the form of $(j,m)$-th KK mass is $m_{(j,m)} = \sqrt{j(j+1)}/R$ and that of the first\,(lightest) KK mass is $M_\text{KK} = \sqrt{2}/R$. There are $2j+1$ numbers of degenerated states in each $j$-level. The exact form of the summation is as follows:
\al{
\sum_{j=1}^{j_{\text{max}}} (2j+1) \ln\paren{\mu \over \sqrt{\frac{j(j+1)}{2}} M_\text{KK}} & &\text{for} \quad 
2 \leq j(j+1)\leq 2 \paren{\mu \over M_\text{KK}}^2. 
}
When we use the approximation: $j(j+1) \simeq j^2$, $j_{\text{max}}$ and $j_{\text{min}}$ are determined as
\al{
j_{\text{max}} \simeq \sqrt{2} \frac{\mu}{m_{\text{KK}}},\quad
j_{\text{min}} \simeq \sqrt{2}.
}
Let us approximate the summation over $j$ by an integral over $j$:
\al{
\sum_{j=1}^{l_{\text{max}}} (2j+1) \ln\paren{\mu \over \sqrt{\frac{j(j+1)}{2}} M_\text{KK}} 
&\simeq \int_{j_{\text{min}}}^{j_{\text{max}}}\cred{dj\,} (2j) \ln\paren{j_{\text{max}} \over j}
\notag \\
&= \paren{\mu \over M_\text{KK}}^2 {-1} \cred{-2\ln{\mu\over M_\text{KK}}}.
}
Thereby we obtain the final form: 
\al{
\alpha_{4i}^{\cred{-1}}(\mu)
	\ \cred{\simeq} \ \alpha_{4i}^{\cred{-1}}(m_Z)
		-\frac{\textsf{b}_i^{\text{SM}}}{2\pi} \ln\paren{\mu \over m_Z}
		-\frac{\textsf{b}_i^{\text{6D}}}{2\pi} \cdot 1
			\sqbr{
				\paren{\mu\over M_\text{KK}}^2	-1
				\cred{	{}
					-2\ln{\mu\over M_\text{KK}}	}
				}.
	\label{RGE_for_S2}
} 
\end{description}
Combining Eq.~\eqref{RGE_for_T2} and Eq.~\eqref{RGE_for_S2}, we get the final form \skipped~\eqref{6DcouplingRGequation}.
%
%
Neglecting the logarithmic terms in Eq.~\eqref{6DcouplingRGequation}, 
we \cred{obtain}
\al{
\alpha_{4i}^{-1} (\Lambda) \sim \alpha_{4i}^{-1} ({m_Z})
- \frac{C \textsf{b}^{\text{6D}}_i}{\cred{2}\pi}
\frac{\Lambda^2 \skipped}{M_\text{KK}^2}.
\label{approximatedalpha4i}
}
\cred{
We note that the coefficient of the quadratic term for $T^2$ coincide with that in Ref.~\cite{Dienes:1998vg} obtained from a different regularization scheme.
}
Putting Eq.~\eqref{approximatedalpha4i} into the condition~$\epsilon(\Lambda)\sim1$, we get
\al{
\Lambda^2
	\sim	\cred{
				{4\pi M_\text{KK}^2
				\over
				C\paren{N_g+2\textsf{b}_i^\text{6D}}\alpha_{4i}(m_Z)},
				}
\label{positionofcutoffscale}
}
\cred{
where we have used $V_2=8\pi C/M_\text{KK}^2$.
Concretely, we get
\al{
\Lambda
	\lesssim
		\begin{cases}
		5.3\,M_\text{KK}	& \text{for $T^2$,}\\
		6.6\,M_\text{KK}	& \text{for $S^2$,}
		\end{cases}
}:
from the $U(1)_Y$ cutoff.
}

In addition to this analysis, we also make consideration for the Landau poles
of the gauge interactions. If the value of the energy where a Landau pole emerges
is smaller than that of the cutoff which we have discussed before,
we should treat the position of the Landau pole: $\alpha_{4i}^{-1} (\Lambda_{\text{Landau}}) = 0$, which is easily obtained with
leading order approximation as
\al{
\Lambda_{\text{Landau}}^{ 2} \sim \frac{\cred{2}\pi M_\text{KK}^2 }{C \textsf{b}^{\text{6D}}_i \alpha_{4i}({m_Z})},
\label{positionofLandaupole}
}
as a cutoff scale instead.
\skipped
The concrete forms of each value are show {in} Table\skipped~\ref{cutoffvaluesT2andS2}.

\begin{table}[t]
\cred{
\begin{center}
\begin{tabular}{|c||c|c|}
\hline 
			&  \multicolumn{2}{|c|}{Type of geometry} \\
\hline
types &  $T^2$-based & $S^2$-based \\
\hline 
$SU(3)_C$ cutoff & no cutoff & no cutoff \\
$SU(2)_W$ cutoff & $6.9\,M_\text{KK}$ & $8.6\,M_\text{KK}$ \\
$U(1)_Y$ cutoff & $5.3\,M_\text{KK}$ & $6.6\,M_\text{KK}$ \\
\hline
$SU(3)_C$ Landau pole & no cutoff & no cutoff \\
$SU(2)_W$ Landau pole & $8.9\,M_\text{KK}$ & $11\,M_\text{KK}$ \\
$U(1)_Y$ Landau pole & $5.4\,M_\text{KK}$ & $6.8\,M_\text{KK}$ \\
\hline
\end{tabular}
\caption{The values of the cutoff scales and the positions of the Landau poles in $T^2$ and $S^2$ cases.}
\label{cutoffvaluesT2andS2}
\end{center}
}
\end{table}%

In the analysis above, we have taken values of $N_g$ as $3$, $2$ and $1$
in each case of $SU(3)_C$, $SU(2)_W$ and $U(1)_Y$, respectively, and
have employed the values 
\begin{equation}
\left\{
\begin{array}{rcl}
\alpha_{U(1)_Y}^{\cred{-1}}(m_Z)  &=& 97.9, \\
\alpha_{SU(2)_W}^{\cred{-1}}(m_Z) &=& 29.4, \\
\alpha_{SU(3)_C}^{\cred{-1}}(m_Z) &=& 8.44,
\end{array}
\right. 
\end{equation}
at $m_Z = 91.1\GeV$.
We do not consider a TeV-scale gauge coupling unification condition as a UV cutoff in this paper.

In the both $T^2$ and $S^2$ cases, the most stringent bounds come from the $U(1)_Y$ cutoff scales, which restrict the effective range of the perturbation the most severely. It is natural that the scale emerging the $U(1)_Y$ Landau pole is near the upper limit of the perturbativity but a little bit higher. 
\skipped

\section{Breit-Wignar formula}
\label{Two_point_function_and_width}
We review how the Breit-Wigner formula emerges from the resummation of the one-particle irreducible (1PI) Higgs two-point function, in order to be careful of possible systematic errors when the width becomes broad. In this paper, we assume that the Higgs mass is larger than twice the $W$ KK mass so that it is sufficient to limit ourselves to the SM case when discussing the Higgs total width. It is straightforward to extend the result for the UED when one wants to take the KK loops into account. 

In the SM, the Higgs production cross section via the gluon fusion process is obtained as
\al{
\hat\sigma_{gg\to H}
	&=	{\pi^2\over8M_H}\,
			\Gamma_{H\to gg}(M_H)\,
			\delta(\shat-M_H^2),
			\label{total_cross_section}
}
where
\al{
\Gamma_{H\to gg}(M_H)
	&=	{\alpha_s^2\over8\pi^3}{M_H^3\over v_\text{EW}^2}\,\ab{I\fn{m_t^2\over M_H^2}}^2.
}
Then we get
\al{
\hat\sigma_{gg\to H\to ZZ}
	&\simeq
		\hat\sigma_{gg\to H}\,
		\Br_{H\to ZZ}(M_H)\nn
	&=	{\pi^2\over8M_H}\Gamma_{H\to gg}(M_H)
		\Br_{H\to ZZ}(M_H)\,
		\delta(\shat-M_H^2),
		\label{narrow_width_limit}
}
where
\al{
\Br_{H\to ZZ}(M_H)
	&=	{\Gamma_{H\to ZZ}(M_H)\over\Gamma_H(M_H)}
	=	{{M_H^3\over32\pi v_\text{EW}^2}
			\sqbr{1-{4m_Z^2\over M_H^2}+{12m_Z^4\over M_H^4}}\sqrt{1-{4m_Z^2\over M_H^2}}\over\Gamma_H(M_H)}.
}
The expression~\eqref{total_cross_section} is obtained in the limit of the vanishing decay width $\Gamma_H\to 0$.
We may introduce a narrow width in Eq.~\eqref{narrow_width_limit} by the Breit-Wigner type replacement
\al{
\delta(\shat-M_H^2)
	&\to	{1\over\pi}{\Delta\over\paren{\shat-M_H^2}^2+\Delta^2}
		\label{delta_replace}
}
to get
\al{
\hat\sigma_{gg\to H\to ZZ}
	&\simeq
		{\pi\over8M_H}
		{\Gamma_{H\to gg}(M_H)\,\Gamma_{H\to ZZ}(M_H)\over\Gamma_H(M_H)}
		{\Delta\over\paren{\shat-M_H^2}^2+\Delta^2}.
		\label{SM_truncated_cross_section}
}
When we want to reproduce the delta function, $\Delta$ in Eq.~\eqref{delta_replace} cannot depend on $\shat$, otherwise we cannot get the correct normalization: $\int d\shat\,\delta(\shat-M_H^2)=1$. One should then perform the replacement of the delta function~\eqref{delta_replace} as
\al{
\Delta\to M_H\Gamma_H(M_H)
	\label{naive_replacement}
}
in Eqs.~\eqref{total_cross_section} and \eqref{narrow_width_limit}.
In literature, see e.g.~\cite{Djouadi:2005gi}, $\Delta$ in Eq.~\eqref{SM_truncated_cross_section} is sometimes replaced as
\al{
\Delta	&\to	{\shat\over M_H}\Gamma_H(M_H).
	\label{Djouadi_replacement}
}

Instead of the truncation~\eqref{narrow_width_limit}, we have already obtained the full $gg\to H\to ZZ$ cross section~\eqref{SM_gg_H_ZZ} by the naive Breit-Wignar type replacement $\Delta\to m_H\Gamma_H$ in and only in the denominator of the Higgs propagator
\al{
{i\over Q^2-M_H^2+i\Delta}.
}
Hereafter, let us see that this treatment gives sufficiently good approximation to the full result~\eqref{SM_gg_H_ZZ_with_Pi}.

\subsection{Resummed propagator}
First let us review how the resummed propagator is obtained in the SM. We write the bare Higgs mass and field in terms of the renormalized ones and the counter terms
\al{
M_B^2
	&=	M_H^2+\delta M_H^2,	&
H_B	&=	\sqrt{Z_H}H,	&
Z_H	&=	1+\delta Z_H.
}
The resummed bare propagator reads
\al{
D_B
	&=	{i\over Q^2-M_B^2+\Pi_H(Q^2)},
		\label{bare_propagator}
}
where $M_B$ and $\Pi_H(Q^2)$ are the bare mass and the 1PI two-point function, respectively, both of which contain ultraviolet (UV) divergences, and $Q^2:=-q^2$ for a Higgs four-momentum $q$.
The renormalized propagator then becomes
\al{
D_R
	&=	{D_B\over Z_H}
	=	{i\over Q^2-M_H^2+\hat\Sigma_H(Q^2)},
		\label{renormalized_propagator}
}
where
\al{
\hat\Sigma_{\cred{H}}(Q^2)
	&:=	\Sigma_H(Q^2)-Z_H\,\delta M_H^2+\delta Z_H\paren{Q^2-M_H^2}
}
is the renormalized (finite) 1PI two-point function, with $\Sigma_H(Q^2):=Z_H\,\Pi_H(Q^2)$ being the 1PI two-point function that is given in terms of the renormalized fields but still contains UV divergences. 
Note that it is sufficient to consider the case $Q^2>0$ for our purpose since we have only $s$-channel Higgs propagator, though $Q^2$ can be negative when the Higgs is virtual, e.g.\ when exchanged in $t$-channel. The Higgs two point function in the SM is given by~\cite{Denner:1991kt}
\al{
\Sigma_H(Q^2)
	&=	-{1\over16\pi^2v^2}\Bigg\{
			{6m_t^2}
				\sqbr{2A_0(m_t^2)+(4m_t^2-Q^2)B_0(Q^2,m_t^2)}\nn
	&\phantom{=	-{1\over16\pi^2v^2}\Bigg\{}
			-2
				\sqbr{
					\paren{6m_W^4-2Q^2m_W^2+{M_H^4\over2}}B_0(Q^2,m_W^2)
					+\paren{3m_W^2+{M_H^2\over2}}A_0(m_W^2)
					-6m_W^4
					}\nn
	&\phantom{=	-{1\over16\pi^2v^2}\Bigg\{}
			-
				\sqbr{
					\paren{6m_Z^4-2Q^2m_Z^2+{M_H^4\over2}}B_0(Q^2,m_Z^2)
					+\paren{3m_Z^2+{M_H^2\over2}}A_0(m_Z^2)
					-6m_Z^4
					}\nn
	&\phantom{=	-{1\over16\pi^2v^2}\Bigg\{}
			-{3\over2}\sqbr{
					{3M_H^4}B_0(Q^2,M_H^2)
					+{M_H^2}A_0(M_H^2)
					}
			\Bigg\},
}
where 
the loop functions are
\al{
A_0(m^2)
	&=	m^2\paren{\Delta-\ln{m^2\over\mu^2}+1},
}
with $\Delta:={2\over\eps}-\gamma+\ln4\pi$ for $D=4-\eps$, and
\al{
B_0(Q^2,m^2)
	&=	\Delta-\int_0^1dx\ln{Q^2x(x-1)+m^2-i\vep\over\mu^2}\nn
	&=	\Delta-\ln{m^2\over\mu^2}+\mc I\fn{4m^2\over Q^2},
}
where
\al{
\mc I(\kappa)
	&:=	
		\int_0^1dx{x\paren{2x-1}
				\over
					\paren{x-{1\over2}}^2-{1-\kappa\over4}-i\vep
					}
	=
		\int_0^1dx{2x^2-x
				\over
					x^2-x+{\kappa\over4}-i\vep
					}.
}
For concreteness we write down
\al{
\mc I(\kappa)
	&=	\begin{cases}
			2\paren{
			1
			-\sqrt{\kappa-1}\arctan\paren{1\over\sqrt{\kappa-1}}
			}	&	\text{($\kappa\geq1$ or $\im\kappa\neq0$),}\\
			2-\sqrt{1-\kappa}\,\ln\paren{1+\sqrt{1-\kappa}\over1-\sqrt{1-\kappa}}
				+i\pi\sqrt{1-\kappa}	&	(0<\kappa\leq 1),
		\end{cases}
		\label{I_kappa_eq}\\
\mc I'(\kappa)
	&=	\begin{cases}
		{1\over\kappa}
			-{\arctan\fn{1\over\sqrt{\kappa-1}}\over\sqrt{\kappa-1}}
				&	\text{($\kappa>1$ or $\im\kappa\neq0$),}\\
		{1\over\kappa}
			+{1\over2\sqrt{1-\kappa}}\ln\paren{1+\sqrt{1-\kappa}\over1-\sqrt{1-\kappa}}
			-{i\pi\over2\sqrt{1-\kappa}}
				&	(0<\kappa<1).
		\end{cases} 
}
Note that when $\im\kappa=0$,
\al{
\im \mc I(\kappa)
	&=	\pi\sqrt{1-\kappa}\,\theta(1-\kappa),	&
\im \mc I'(\kappa)
	&=	-{\pi\over2\sqrt{1-\kappa}}\,\theta(1-\kappa).
}
In particular,
\al{
\mc I(4)	&=	{6-\sqrt{3}\pi\over3}
		\simeq
			0.18,	&
\mc I'(4)	&=	-{2\sqrt{3}\pi-9\over36}
		\simeq
			-0.052,
}
and for small and large $\kappa$,
\al{
\mc I(\kappa)
	&=	2+\ln{\kappa\over4}+i\pi+O\fn{\kappa\ln\kappa}&
	&(\kappa\ll1),\\
\mc I(\kappa)
	&=	{2\over3\kappa}+O\fn{\kappa^{-2}}&
	&(\kappa\gg1).
}
\subsection{On-shell scheme renormalization}
In the on-shell scheme, we put the following renormalization conditions:
\al{
\left.\re\hat\Sigma_H(Q^2)\right|_{Q^2=M_H^2}
	&=	0,	&
\left.\re{\partial\over\partial Q^2}\hat\Sigma_H(Q^2)\right|_{Q^2=M_H^2}
	&=	0,
		\label{on_shell_renormalization_condition}
}
which gives
\al{
\hat\Sigma_H(Q^2)
	&=	\Sigma_H(Q^2)-\re\Sigma_H(M_H^2)-\re\Sigma_H'(M_H^2)\paren{Q^2-M_H^2}\nn
	&=	-{1\over16\pi^2 v^2}\Bigg(
			6m_t^2\Bigg\{
				\paren{4m_t^2-Q^2}
					\sqbr{\mc I\fn{4m_t^2\over Q^2}-\re\mc I\fn{4m_t^2\over M_H^2}}\nn
	&\phantom{=	-{1\over16\pi^2 v^2}\Bigg(
			6m_t^2\Bigg\{}
				+\paren{4m_t^2-M_H^2}{4m_t^2\over M_H^4}\re \mc I'\fn{4m_t^2\over M_H^2}
					\paren{Q^2-M_H^2}
				\Bigg\}\nn
	&\phantom{=	-{1\over16\pi v^2}\Bigg(}
		-2\Bigg\{
				\paren{6m_W^4-2Q^2m_W^2+{M_H^4\over2}}
					\sqbr{\mc I\fn{4m_W^2\over Q^2}-\re \mc I\fn{4m_W^2\over M_H^2}}\nn
	&\phantom{=	-{1\over16\pi v^2}\Bigg(-2\Bigg\{}
				+\paren{6m_W^4-2M_H^2m_W^2+{M_H^4\over2}}
				{4m_W^2\over M_H^4}\re \mc I'\fn{4m_W^2\over M_H^2}\paren{Q^2-M_H^2}\Bigg\}\nn
	&\phantom{=	-{1\over16\pi v^2}\Bigg(}
		-\Bigg\{
				\paren{6m_Z^4-2Q^2m_Z^2+{M_H^4\over2}}
					\sqbr{\mc I\fn{4m_Z^2\over Q^2}-\re \mc I\fn{4m_Z^2\over M_H^2}}\nn
	&\phantom{=	-{1\over16\pi v^2}\Bigg(-\Bigg\{}
				+\paren{6m_Z^4-2M_H^2m_Z^2+{M_H^4\over2}}
				{4m_Z^2\over M_H^4}\re \mc I'\fn{4m_Z^2\over M_H^2}\paren{Q^2-M_H^2}\Bigg\}\nn
	&\phantom{=	-{1\over16\pi v^2}\Bigg(}
		-{9M_H^4\over2}\Bigg\{
			\sqbr{\mc I\fn{4M_H^2\over Q^2}-\mc I\fn{4}}
			+{4\over M_H^2}\mc I'\fn{4}\paren{Q^2-M_H^2}
			\Bigg\}
		\Bigg),
}
and hence
\al{
\im\hat\Sigma_H(Q^2)
	&=	-{1\over16\pi^2 v^2}\Bigg\{
			-6m_t^2Q^2
					\pi\sqbr{1-{4m_t^2\over Q^2}}^{3/2}\,\theta\fn{1-{4m_t^2\over Q^2}}\nn
	&\phantom{=	-{1\over16\pi v^2}\Bigg(}
		-M_H^4\paren{{12m_W^4\over M_H^4}-{4Q^2m_W^2\over M_H^4}+1}
					\pi\sqrt{1-{4m_W^2\over Q^2}}\,\theta\fn{1-{4m_W^2\over Q^2}}\nn
	&\phantom{=	-{1\over16\pi v^2}\Bigg(}
		-{M_H^4\over2}\paren{{12m_Z^4\over M_H^4}-{4Q^2m_Z^2\over M_H^4}+1}
					\pi\sqrt{1-{4m_Z^2\over Q^2}}\,\theta\fn{1-{4m_Z^2\over Q^2}}\nn
	&\phantom{=	-{1\over16\pi v^2}\Bigg(}
		-{9M_H^4\over2}\pi\sqrt{1-{4M_H^2\over Q^2}}\,\theta\fn{1-{4M_H^2\over Q^2}}
		\Bigg\}.
		\label{ImSigma}
}
We can compare this result with the tree-level Higgs decay width
\al{
\Gamma_H^\text{tree}(M_H)
	&=	{M_H^3\over16\pi v_\text{EW}^2}
			\sqbr{1-{4m_W^2\over M_H^2}+{12m_W^4\over M_H^4}}\sqrt{1-{4m_W^2\over M_H^2}}\,\theta\fn{M_H-2m_W}\nn
	&\quad
		+{M_H^3\over32\pi v_\text{EW}^2}
			\sqbr{1-{4m_Z^2\over M_H^2}+{12m_Z^4\over M_H^4}}\sqrt{1-{4m_Z^2\over M_H^2}}\,\theta\fn{M_H-2m_Z}\nn
	&\quad
		+{3M_Hm_t^2\over8\pi v_\text{EW}^2}
			\sqbr{1-{4m_t^2\over M_H^2}}^{3/2}\theta\fn{M_H-2m_t}.
			\label{tree_width}
}
We see from Eq.~\eqref{ImSigma} that the leading term for the $Q\sim M_H\gg 2m_t$ limit is not proportional to $Q^2$ for the $W$ and $Z$ contributions and therefore the replacement~\eqref{Djouadi_replacement}, namely
\al{
D_R
	\to	{i\over Q^2-M_H^2+i{Q^2\over M_H}\Gamma_H}
}
does not give a good fit.

When computing the full cross section for the process $gg\to H\to ZZ$, we may employ the resummed propagator~\eqref{renormalized_propagator}.
Neglecting the contributions from box diagrams, we get the cross section
\al{
\hat\sigma_{gg\to H\to ZZ}
	&=	{\alpha_s^2\over256\pi^2}\paren{m_Z\over v_\text{EW}}^4
		\sqbr{
			1+{\paren{\shat-2m_Z^2}^2\over 8m_Z^4}
			}
		\sqrt{1-{4m_Z^2\over\shat}}\,
		\ab{I\fn{m_t^2\over\shat}}^2\nn
	&\quad
		\times{1\over\pi}{\shat\over\paren{\shat-m_H^2+\re\hat\Sigma_H\fn{\shat}}^2+\paren{\im\hat\Sigma_H(\shat)}^2}.
		\label{SM_gg_H_ZZ_with_Pi}
}

In Fig.~\ref{various_widths}, we show the results for various replacement. We see that any replacement suffices when Higgs mass is not very large, $M_H=300\GeV$ (left). (This is the case for the Higgs mass above the top threshold $m_H\gtrsim 2m_t$ too.) For the large Higgs mass (right), the decay width becomes larger and we see that our approximation~\eqref{SM_gg_H_ZZ} with Breit-Wignar type replacement in the denominator $\Delta\to M_H\Gamma_H$ gives a good fit to the cross section with full two-point function~\eqref{SM_gg_H_ZZ_with_Pi}.
\begin{figure}[t]
\begin{center}
		\mbox{
			\includegraphics[width=.4\textwidth, bb= 0 0 488 293, clip]{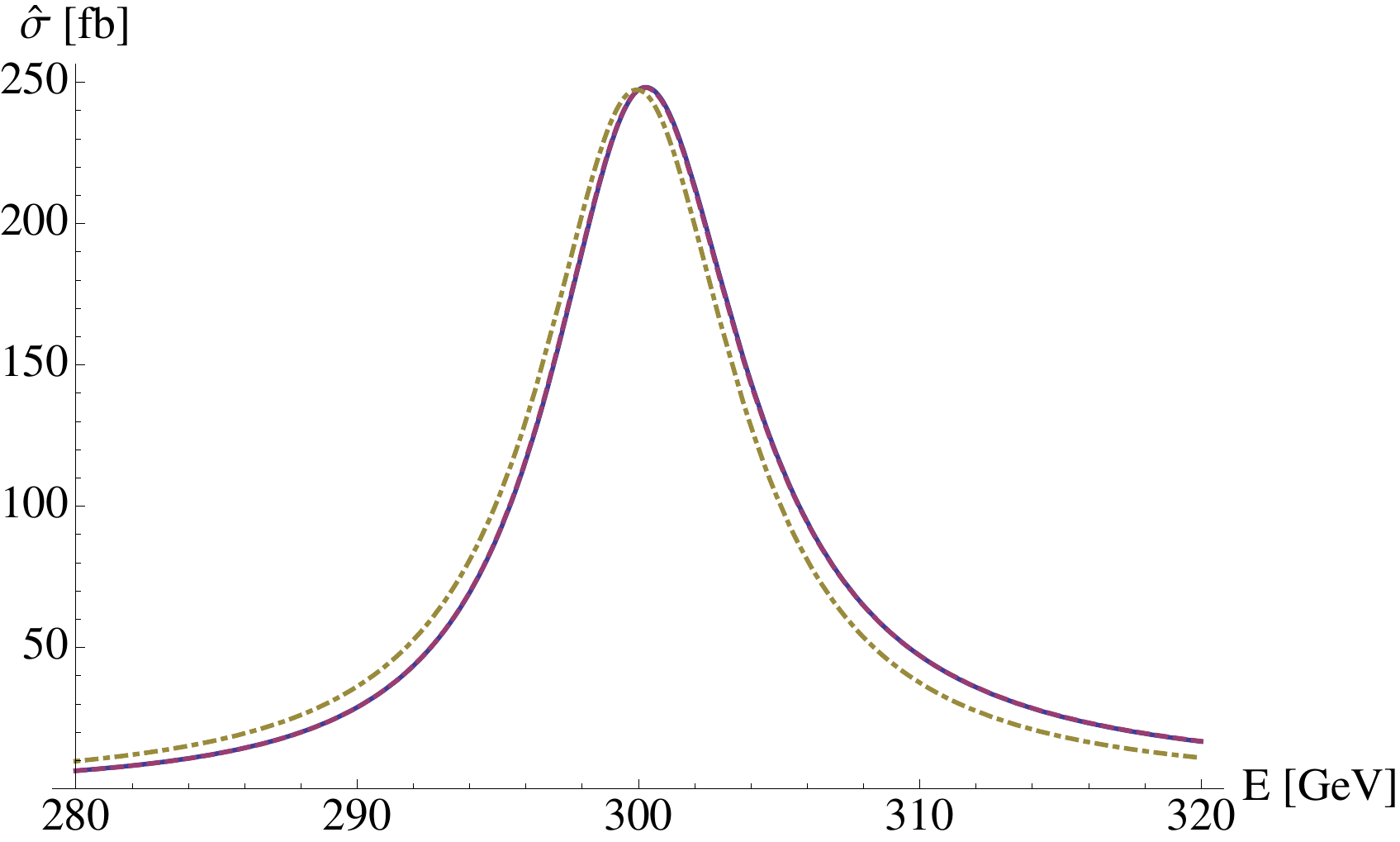}
			\includegraphics[width=.4\textwidth, bb= 0 0 435 264,clip]{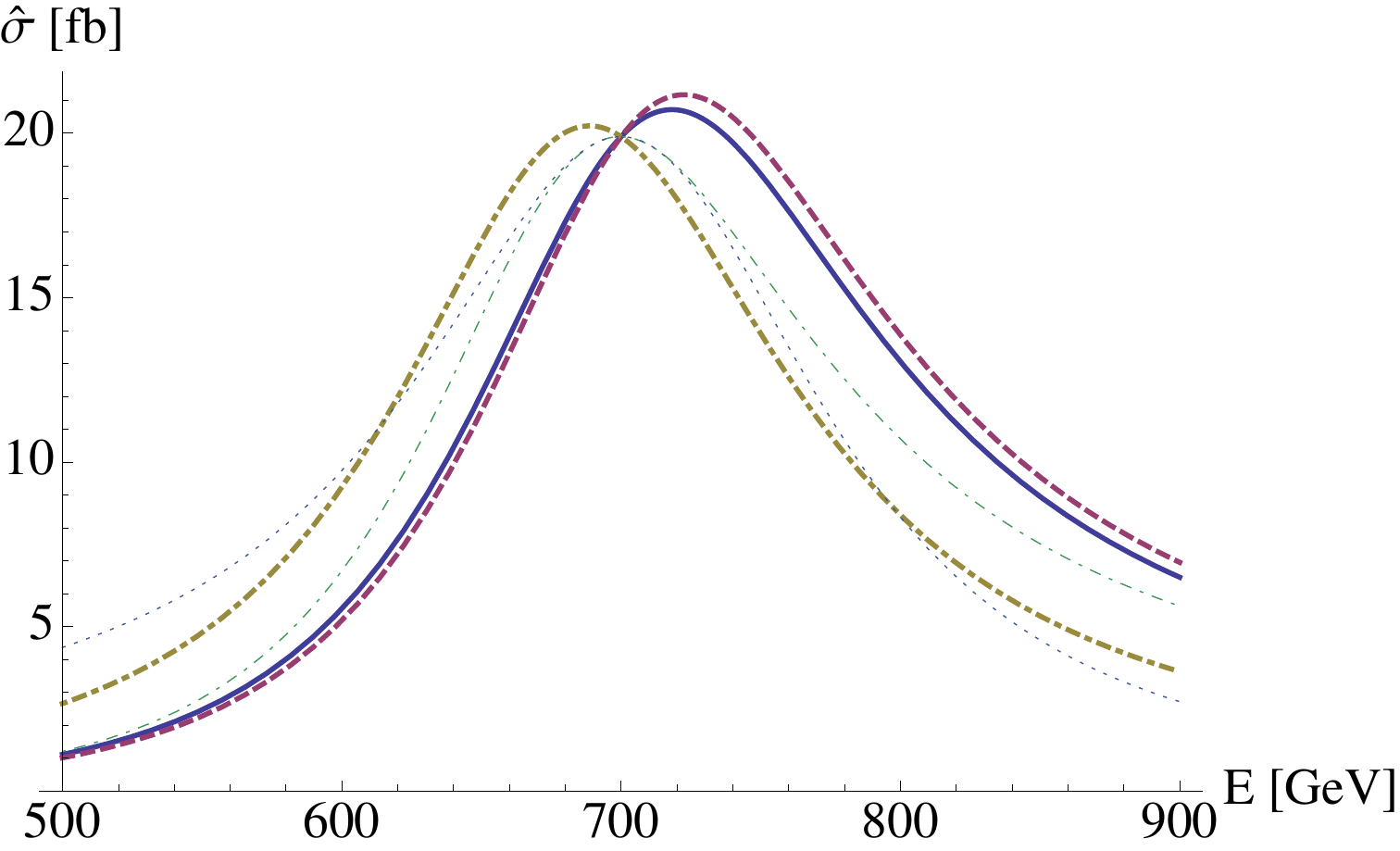}
			}
\caption{
Parton level cross section for the process $gg\to H\to ZZ$ with $m_H=300\,\text{GeV}$ (left) and $700\,\text{GeV}$ (right). Inclusion of full two point function~\eqref{SM_gg_H_ZZ_with_Pi} (blue solid line), the full one~\eqref{SM_gg_H_ZZ} with $\Delta\to M_H\Gamma_H$ (red dashed), and the truncated one~\eqref{SM_truncated_cross_section} with $\Delta\to\shat\Gamma_H/M_H$ (yellow dot-dashed) are shown. In the right figure, the full one~\eqref{SM_gg_H_ZZ} with $\Delta\to\shat\Gamma_H/M_H$ (thin dot-dashed) and the truncated one~\eqref{SM_truncated_cross_section} with $\Delta\to M_H\Gamma_H$ (thin dotted) are also drawn.
}
\label{various_widths}
\end{center}
\end{figure}

Finally, just for comparison with Eq.~\eqref{SM_truncated_cross_section}, let us present the resummed cross section~\eqref{SM_gg_H_ZZ_with_Pi} in a rewritten form
\al{
\hat\sigma_{gg\to H\to ZZ}
	&=	{\pi\over8M_H}\,
		{\alpha_s^2M_H^3\over 8\pi^3 v_\text{EW}^2}
		\ab{{\shat\over M_H^2}I\fn{m_t^2\over\shat}}^2\nn
	&\quad
		\times
		{{\shat\over M_H}\Gamma_H(M_H)\over\paren{\shat-M_H^2+\re\Pi_H(\shat)}^2+\paren{\im\Pi_H(\shat)}^2}\nn
	&\quad
		\times{{M_H^3\over32\pi v_\text{EW}^2}
			\sqbr{
				1-{4m_Z^2\over\shat}+{12m_Z^4\over\shat^2}
				}
			\sqrt{1-{4m_Z^2\over\shat}}
			\over
			\Gamma_H(M_H)
			}.
}

\subsection{Pole scheme renormalization}

\begin{figure}[t]
\begin{center}
			{
			\hfill
			\includegraphics[width=.4\textwidth, bb= 0 0 288 178, clip]{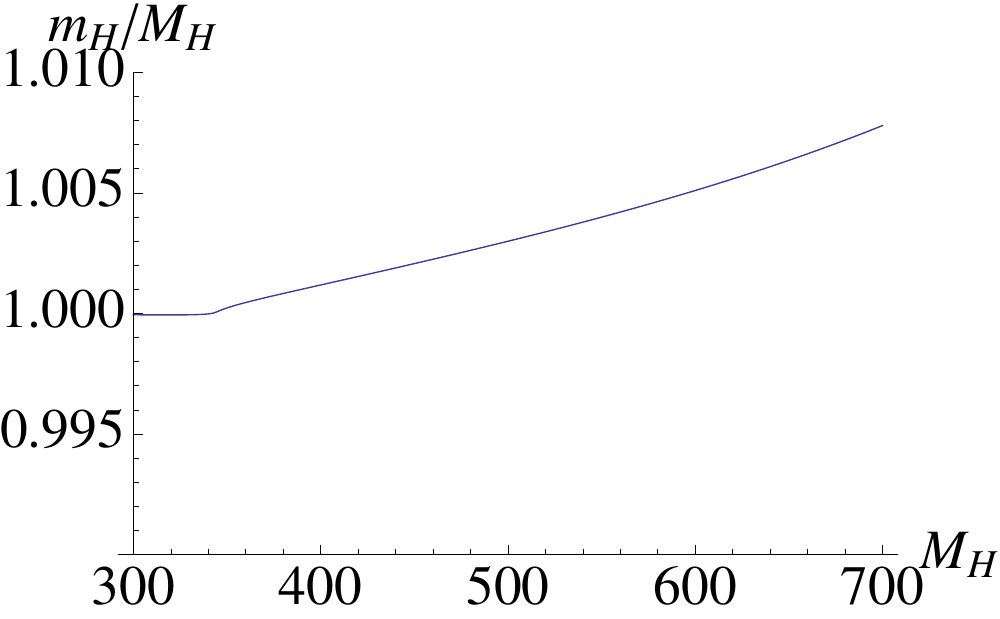}
			\hfill
			\includegraphics[width=.4\textwidth, bb= 0 0 288 165,clip]{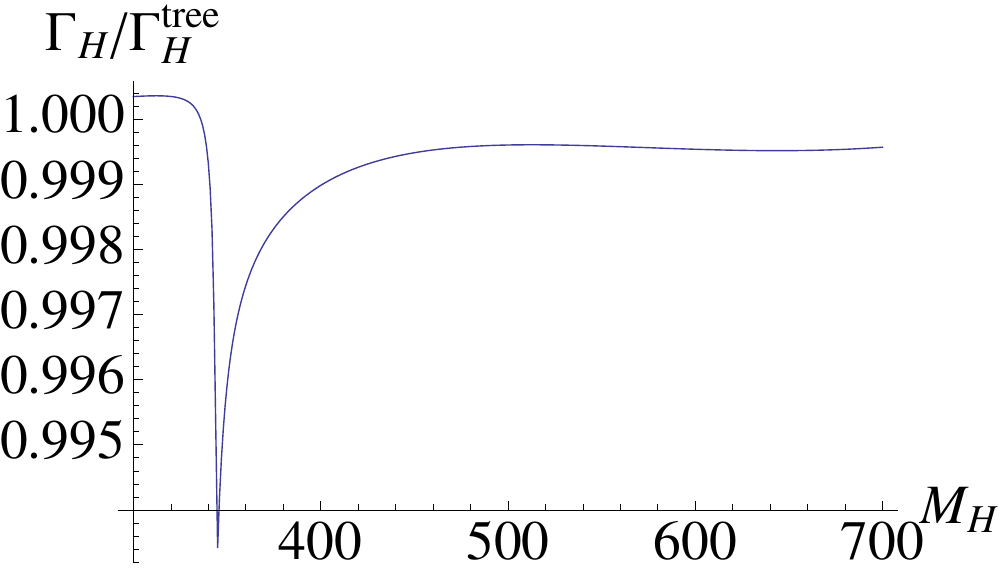}
			}
\caption{Ratio of the pole mass $m_H$ to the on-shell one $M_H$ as a function of the on-shell mass (left); same for the ratio of the pole-scheme width $\Gamma_H$ to the tree-level one $\Gamma_H^\text{tree}$ (right). \label{ratios_vs_MH}
}
\end{center}
\end{figure}

Let us review the pole scheme renormalization~\cite{Kniehl:2001ch}. Instead of the on-shell condition~\eqref{on_shell_renormalization_condition}, the pole scheme renormalization condition fixes the pole of the \skipped propagator~\eqref{bare_propagator} at $\ol{Q^2}$ which we parametrize by two real constants $m_H$ and $\Gamma_H$ as: $\ol{Q^2}=m_H^2-im_H\Gamma_H$, namely,
\al{
\ol{Q^2}-M_B^2+\Pi_H(\ol{Q^2})
	&=	0.
}
As the pole position of the bare propagator~\eqref{bare_propagator} is the same as that of the renormalized one~\eqref{renormalized_propagator}, we see that the on-shell renormalized two-point function satisfies
\al{
\ol{Q^2}-M_H^2+\hat\Sigma_H(\ol{Q^2})
	=	0,
}
that is,
\al{
M_H^2
	&=	m_H^2+\re\hat\Sigma_H(\ol{Q^2}),	&
m_H\Gamma_H
	&=	\im\hat\Sigma_H(\ol{Q^2}).
		\label{complex_mass_and_width}
}
We see that the real and imaginary parts determine the pole scheme mass $m_H$ and decay rate $\Gamma_H$ as functions of the on-shell scheme mass $M_H$. Note that the renormalized two-point function~$\hat\Sigma_H$ has implicit dependence on the on-shell mass $M_H$.

In Fig~\ref{ratios_vs_MH}, we plot the ratio of the pole to on-shell mass $m_H/M_H$ (left) and the pole-scheme width $\Gamma_H$ to the tree-level width~\eqref{tree_width} (right) as functions of the on-shell mass $M_H$, computed within the SM. 

It is known that the on-shell scheme mass is gauge dependent at the NNLO, see references in~\cite{Kniehl:2001ch}. In contrast, the pole position of the amplitude is a gauge independent physical notion. Therefore, in principle we should utilize the pole scheme mass and width. \cred{However, the on-shell mass and the pole mass are identical at consistent 1-loop order (which is the highest order considered here), and they start differing only at 2-loop order and above.}
\cmagenta{We see from Fig.~\ref{ratios_vs_MH} that the both schemes agree within 1\% accuracy in our approximated treatment neglecting non-resonant box contributions, which are the same order as $O(\alpha)$-corrections to the resonant ones.}\footnote{\cmagenta{The difference in Fig.~\ref{ratios_vs_MH} stems from the fact that a complex value is inserted for $\overline{Q^2}$ in Eq.~\eqref{complex_mass_and_width}, i.e.\ $\overline{Q^2}=m_H^2 - i m_H \Gamma_H$. However, $\Gamma_H$ is formally a higher-order term, since it corresponds to the imaginary part of the self-energy, which first occurs at one-loop order. Therefore, when inserted into the one-loop self-energies in Eq.~\eqref{complex_mass_and_width}, this leads to a contribution that is formally at the NNLO level, which causes the small numerical difference between the two schemes in Fig.~\ref{ratios_vs_MH}.}}
\cred{We} safely utilize the on-shell mass and the tree-level width~\eqref{tree_width} even though the Higgs decay width becomes as large as 180\,GeV when $M_H=700\,\text{GeV}$.

\cmagenta{
Note that our treatment to include the decay width in the numerical calculation~\eqref{SM_gg_H_ZZ} with~\eqref{Delta_replacement} corresponds to the (gauge invariant) complex-mass scheme, see e.g.\ Refs.~\cite{Denner:1999gp,Denner:2005fg}, where the real Higgs mass is replaced by a complex value everywhere in the amplitude, which in our case leads to a replacement of the Higgs mass in its propagator only.
}




\bibliographystyle{TitleAndArxiv}
\bibliography{paper}

\end{document}